\documentclass[aps,prd,10pt,notitlepage,superscriptaddress,nofootinbib,numbers]{revtex4-2}

\usepackage[utf8]{inputenc}
\usepackage[T1]{fontenc}
\usepackage{lmodern}
\usepackage{microtype}
\usepackage{amsmath,amssymb,amsfonts,mathrsfs}
\usepackage{bm}
\usepackage{booktabs}
\usepackage{graphicx}
\usepackage[dvipsnames]{xcolor}
\usepackage{hyperref}
\hypersetup{colorlinks=true,citecolor=Purple,linkcolor=Purple,urlcolor=Purple,breaklinks=true}
\usepackage[capitalize]{cleveref}
\usepackage{tikz}
\usepackage{lipsum}

\definecolor{lime}{HTML}{A6CE39}
\DeclareRobustCommand{\orcidicon}{%
	\begin{tikzpicture}
		\draw[lime, fill=lime] (0,0)
		circle [radius=0.16]
		node[white] {{\fontfamily{qag}\selectfont \tiny ID}};
		\draw[white, fill=white] (-0.0625,0.095)
		circle [radius=0.007];
	\end{tikzpicture}
	\hspace{-2mm}
}

\foreach \x in {A, ..., Z}{%
	\expandafter\xdef\csname orcid\x\endcsname{\noexpand\href{https://orcid.org/\csname orcidauthor\x\endcsname}{\noexpand\orcidicon}}
}

\newcommand\orcidJonathan{{\href{https://orcid.org/0000-0001-9291-0893}{\orcidicon}}}
\newcommand\orcidEdson{{\href{https://orcid.org/0000-0001-9929-5977}{\orcidicon}}}

\newcommand{\dd}{\mathrm{d}}
\newcommand{\Mone}{\mathcal{M}_1}
\newcommand{\Mtwo}{\mathcal{M}_2}
\newcommand{\Popt}{\mathcal{P}}
\newcommand{\Iobs}{I_{\mathrm{obs}}}
\newcommand{\Iem}{I_{\mathrm{em}}}

\begin{document}

\title{Optical effects of one-sided Kiselev-type quintessence in reflection-asymmetric polymer thin-shell wormholes}

\author{Jonathan A. Rebou\c{c}as\orcidJonathan\!\!}
\email{jalvesreboucas@ifce.edu.br}
\affiliation{Instituto Federal de Educa\c{c}\~ao, Ci\^encia e Tecnologia do Cear\'a (IFCE), Iguatu-CE, Brazil}

\author{Edson Otoniel\orcidEdson\!\!}
\email{edson.otoniel@ufca.edu.br}
\affiliation{Universidade Federal do Cariri (UFCA), Instituto de Forma\c{c}\~ao de Educadores - IFE,  R. Oleg\'ario Emidio de Araujo S/N, Brejo Santo - CE, 63.260-000 - Brazil}

\date{\today}

\begin{abstract}
Strong-field rays that traverse a wormhole can carry optical information about an exterior inaccessible to the observer. We investigate this mechanism in a reflection-asymmetric thin-shell wormhole formed by joining a pure-polymer exterior containing the observer and emitting disk to a polymer geometry with a Kiselev-type anisotropic environment on the opposite side. Equal masses and a common polymer scale, angular geometry, and throat position isolate the effect of one-sided quintessence. We derive the shell-frame optical matching and calculate matched critical scales, null-ray topologies, orbit numbers, transfer functions, and face-on intensity maps for three emission prescriptions. The matching projects the opposite-side photon barrier onto a distinct inner edge of the observer screen, while the outer cosmological-type horizon partitions cross-throat rays into returned and terminal families. Across the admissible parameter grid, increasing the environmental amplitude widens the returned screen interval, increasing the polymer correction contracts it, and changing the environmental exponent produces competing trends. Observer-side orbit numbers and ordinary transfer branches remain invariant under changes confined beyond the throat, providing an exact control, whereas returned branches move and broaden. Once these branches enter the emitting support, they generate additional inner annuli at source-dependent onset strengths. The combined geometric and radiative analysis identifies a controlled cross-throat optical response to one-sided quintessence and shows how environmental information beyond the throat is encoded in matched critical structure and image morphology.
\end{abstract}


\maketitle

\tableofcontents
\clearpage

\section{Introduction}\label{sec:introduction}

Strong-field images provide a direct way to organize how null propagation, compact-object geometry, and luminous matter combine on an observer's screen.  The boundary between captured and escaping rays was identified in early studies of photon escape, while subsequent disk imaging showed how strongly bent trajectories map an emitting flow into multiple image components \cite{Synge1966,Luminet1979}.  This geometric program acquired observational urgency with proposals to resolve horizon-scale silhouettes and with horizon-scale observations of the supermassive objects in Messier~87 and at the Galactic Center \cite{FalckeMeliaAgol2000,EHTM872019,EHTSgrA2022}.  Nevertheless, an observed brightness distribution is not a metric observable by itself: it also depends on the emitting medium, viewing geometry, absorption, and instrumental response.  Optical calculations are therefore most informative when they distinguish robust propagation structures from source-dependent intensity features.

For static spherical spacetimes, unstable circular photon orbits organize the critical impact parameters and the logarithmic accumulation of strongly deflected rays \cite{Bozza2002,PerlickTsupko2022}.  Their screen imprint must, however, be described with care.  The critical curve is a geometrical separator in ray space, whereas the brightness and width of lensing and photon-ring components depend on how often the rays intersect the source and on the emissivity assigned to those intersections \cite{Gralla2019}.  The narrow substructure generated by near-critical propagation has potentially distinctive interferometric behavior \cite{Johnson2020}, but translating such structure into a claim about the nature of a compact object requires control of astrophysical degeneracies and of the assumptions used to model horizonless alternatives \cite{CardosoPani2019}.

Traversable wormholes provide a particularly useful setting in which to separate these ingredients because their global connectivity differs from that of a black hole even when one exterior is locally similar.  The geometric and matter requirements of static wormhole throats were systematized in the Morris--Thorne framework \cite{MorrisThorne1988}, while the junction conditions for singular hypersurfaces supply the invariant relation between a shell stress tensor and the discontinuity of extrinsic curvature \cite{Israel1966}.  Cut-and-paste surgery then permits two retained exterior regions to be joined at a timelike throat without extending either seed geometry through its original interior \cite{Visser1989}.  The existence of a static junction does not by itself establish radial stability: that is a separate perturbative problem involving the shell response or equation of state \cite{PoissonVisser1995}.  Recent analyses have also clarified that relations between circular photon orbits and shell stability are conditional rather than interchangeable, reinforcing the need to keep optical and dynamical claims logically distinct \cite{TsukamotoKokubu2024}.

Reflection asymmetry adds a second layer to this construction.  When the two retained regions possess different lapse functions or critical null scales, a ray crossing the throat can probe an optical barrier that is absent from the observer's exterior.  Early studies showed that this contralateral structure can produce nonstandard shadow boundaries and double-shadow configurations \cite{Wang2020,Wielgus2020}.  In disk illumination, the same mechanism generates additional photon-ring branches whose positions on the observer's screen are fixed not by a naive identification of impact parameters but by the shell matching of locally measured photon data \cite{Peng2021}.  Separate investigations have examined the radial stability of reflection-asymmetric shells that support double shadows \cite{Tsukamoto2021} and demonstrated that related optical configurations can occur with positive surface energy in other gravitational settings \cite{Guerrero2021}.  These results make the asymmetry physically consequential, but they also show why shell support, stability, and optical appearance must be evaluated as different questions.

The resulting phenomenology is sensitive to the chosen pair of exterior geometries.  Hayward-profile thin shells exhibit transfer-function and image structures tied to their regular seed \cite{Guo2023}; asymmetric Horndeski constructions and Kalb--Ramond backgrounds provide distinct realizations of additional photon rings \cite{Luo2024Horndeski,TanLan2026KalbRamond}; and configurations with more than one relevant critical orbit can develop a multiphoton-ring hierarchy \cite{Macedo2026Multiphoton}.  These examples establish a broad mechanism---cross-throat propagation exposes critical structure from the second exterior---but do not make its quantitative imprint universal.  The observable intervals, the number of returned branches, and their magnification must instead be derived from the actual metric functions, angular sector, shell position, and causal domain of each model.

Polymer black-hole geometries offer a controlled arena for this comparison because quantum-corrected effective dynamics replaces the classical central region while retaining a tractable static exterior.  The polymer Schwarzschild construction used here descends from effective black-to-white-hole geometries and their canonical-variable formulations \cite{Bodendorfer2019Extended,Bodendorfer2021Variables}.  Its global and phenomenological properties have been investigated beyond a single coordinate representation, including generic features of polymer quantum black holes and periodic-orbit dynamics \cite{Munch2023GenericFeatures,Tu2023PeriodicOrbits}.  Of particular importance for optics, the symmetry spheres of the present model are governed by a nonareal angular function, so photon-sphere conditions and disk radii cannot be inferred by replacing that function with the squared radial coordinate.  Pure-polymer exteriors have also been used to construct thin-shell wormholes and study their radial stability \cite{Javed2022}; those results establish useful dynamical context for the seed family and complement the optical analysis developed below.

The second exterior introduces a Kiselev-type surrounding component.  Kiselev's solution provides a widely used static spherical parametrization of an anisotropic matter environment through a normalization and an equation-of-state-like exponent \cite{Kiselev2003}, and its strong-lensing consequences already indicate that this term can shift the critical null structure \cite{Younas2015StrongGravitational}.  The word ``quintessence'' in this context is conventional: the associated stress tensor is anisotropic and should not be identified with a perfect-fluid cosmological quintessence field \cite{Visser2019KiselevBlack}.  When this contribution is combined with the polymer lapse, it modifies both the accessible static domain and the optical potential, as established for the corresponding surrounded polymer black hole \cite{Araujo2025}.  More generally, environmental thin-shell constructions show that surrounding matter can alter junction properties and observable scales without changing the logical separation between the chosen background and the shell itself \cite{Reboucas2026Voids}.

A symmetric thin-shell wormhole assembled from two copies of the polymer--Kiselev exterior has recently been analyzed across its junction, stability, thermodynamic, and optical sectors \cite{Reboucas2026PolymerQuintessence}.  That construction demonstrated the relevance of retaining the nonareal angular geometry and of comparing cross-throat images with a black-hole control, while reflection symmetry makes both exteriors respond simultaneously to the surrounding component.  The present study develops the complementary controlled asymmetric pairing in which the observer and emitting disk remain in a pure-polymer exterior while only the opposite side carries the Kiselev-type term.  Such a pairing isolates which screen features are genuinely transmitted through the throat from those produced locally, and it identifies the outer zero of the second-side lapse as the boundary of the adopted static patch.

In this work, we construct that pairing with equal mass scales, a common polymer parameter, the same nonareal angular function, and a fixed throat, so that the reflection asymmetry is introduced solely by the Kiselev parameters on the side opposite the observer.  We derive the static junction and the shell-frame matching of null data, identify the two matched critical scales on a single observer screen, and classify rays that scatter on the observer side, return after a turning point in the second exterior, or terminate at its outer cosmological-type horizon.  We then compute orbit numbers, transfer functions, and optically thin face-on intensity maps for three phenomenological emission profiles.  The analysis shows how the opposite-side environment relocates the inner matched critical scale and controls the availability of returned higher-order branches, while the observer-side orbit number remains insensitive to parameters that occur only beyond the throat.  The image-level response acquires source-dependent activation thresholds, thereby separating propagation effects from the emissive support and providing a controlled optical measure of the one-sided environment.

This paper is organized as follows.  Section~\ref{sec:geometry} introduces the pure-polymer and polymer--Kiselev exteriors and their causal domains.  Section~\ref{sec:shell} constructs the static asymmetric junction and derives the optical matching conditions.  Section~\ref{sec:null-geodesics} develops the matched null potential and critical structure, Sec.~\ref{sec:ray-topology} classifies the allowed ray topologies, and Sec.~\ref{sec:transfers} presents the orbit numbers and transfer functions.  Section~\ref{sec:emission} specifies the circular-motion scales, emission profiles, and observable intensity maps.  Our conclusions are collected in Sec.~\ref{sec:conclusions}.

\section{Polymer and polymer--quintessence exteriors}
\label{sec:geometry}

The construction joins two members of the same quantum-corrected metric family while activating a Kiselev-type environment only on the side opposite the observer.  This choice produces a controlled reflection asymmetry: the polymer mass and length scales, the angular geometry, the throat, the observer, and the source prescription remain fixed, whereas the causal domain and optical potential of the second exterior change.  It therefore differs both from the pure-polymer thin-shell wormhole studied in Ref.~\cite{Javed2022} and from the reflection-symmetric polymer--quintessence construction of Ref.~\cite{Reboucas2026PolymerQuintessence}.  The former established that a polymer black-hole exterior can support a cut-and-paste throat and that its radial stability depends strongly on the surface equation of state and polymer parameter.  The latter retained the nonareal angular sector throughout the junction and radial perturbation analyses, found finite local stability sectors for the variable Chaplygin closure, and showed that cross-throat propagation can add inner image branches even when the wormhole and its black-hole control share the same exterior critical curve.  The present construction complements those results by isolating the optical response produced when the Kiselev environment is confined to the side opposite the observer.

Both retained regions are static and spherically symmetric,
\begin{equation}
 \dd s_i^2=-F_i(r_i)\dd t_i^2+\frac{\dd r_i^2}{F_i(r_i)}
 +H(r_i)\left(\dd\theta^2+\sin^2\theta\,\dd\phi^2\right),
 \qquad i=1,2.
 \label{eq:metric}
\end{equation}
The coordinate $r_i$ is not an areal radius.  A symmetry sphere has area $4\pi H(r_i)$, so every centrifugal term, shell area, photon-sphere condition, and disk radius must be built from $H$, rather than from $r_i^2$.  We restrict the underlying polymer geometry to the symmetric-bounce sector, for which
\begin{equation}
 H(r)=r^2+\ell^2,
 \qquad
 \ell=(\lambda M)^{1/3}.
 \label{eq:angular-function}
\end{equation}
The corresponding lapse is
\begin{equation}
 F_0(r)=\frac{4\ell^2-2M\sqrt{4\ell^2+r^2}+r^2}{\ell^2+r^2}.
 \label{eq:pure-polymer-lapse}
\end{equation}
Here $M$ fixes the common mass scale, $\lambda$ is the polymer parameter, and $\ell$ is the associated length.  In geometrized units, $M$ and $\ell$ have dimensions of length and $\lambda$ has dimension of length squared.  The source construction uses $0\leq\lambda/M^2\leq1$ for the black-hole branch \cite{Bodendorfer2019Extended,Araujo2025}.  At nonzero $\lambda$, the minimum of $H$ belongs to the quantum-corrected interior of the complete seed geometry, but it is excised from the present optical spacetime because only $r_i\geq a>0$ is retained.

The observer and the thin emitting disk are placed in the pure-polymer exterior $\Mone$, where $F_1=F_0$.  This side is asymptotically flat, with $F_1=1-2M/r+\mathcal{O}(r^{-2})$ and $H=r^2+\mathcal{O}(1)$ at large radius.  Consequently its static Killing time is already normalized at infinity, and the impact parameter $b_1=L/E_1$ will be the physical radial coordinate on the observer screen.  In the complete pure-polymer solution the quantum correction regularizes the central sector and replaces the classical singular behavior by a bounce \cite{Bodendorfer2019Extended,Bodendorfer2021Variables}.  The present shell does not image that core directly: it retains only the exterior interval beyond the event horizon and then truncates it at the throat.  Ref.~\cite{Javed2022} used a polymer black-hole background to construct a reflection-symmetric thin-shell wormhole and found that the quantum parameter materially reorganizes the stability regions of the considered surface fluids, with the variable Chaplygin model providing the broadest stable configurations in their sampled analysis.  Together with the polymer--quintessence stability sectors established in Ref.~\cite{Reboucas2026PolymerQuintessence}, these results provide the dynamical context for using the same geometric family in the present optical construction.

The second exterior $\Mtwo$ is described by the hybrid lapse introduced and analyzed in Ref.~\cite{Araujo2025},
\begin{equation}
 F_2(r)=F_q(r)=F_0(r)-\frac{c_q}{r^{3w_q+1}}.
 \label{eq:quintessence-lapse}
\end{equation}
The parameter $w_q$ controls the radial power of the Kiselev sector and is restricted here to the customary open quintessence interval $-1<w_q<-1/3$; $c_q$ is positive and has dimension $\mathsf{L}^{3w_q+1}$.  The dimensionless combinations used below are $\bar\lambda=\lambda/M^2$ and $\bar c_q=c_q/M^{3w_q+1}$.  The endpoint $w_q=-1$ would produce a cosmological-constant-like term, while $w_q=-1/3$ would give a constant lapse shift; neither endpoint belongs to the adopted domain.  Setting $c_q=0$ recovers the pure-polymer exterior without removing the nonareal angular sector, whereas taking $\lambda=0$ gives the Schwarzschild--Kiselev geometry.

The Kiselev deformation changes more than the local value of the lapse.  For the interval above, its magnitude grows toward large radius and destroys asymptotic flatness, although the leading quintessence contributions to the curvature invariants can still decay there \cite{Araujo2025}.  It also makes the hybrid solution singular at the origin; that singular region is again removed by the shell.  Most importantly for the present problem, the positive-lapse domain is generally finite.  The published parameter study found an inner black-hole-type horizon and an outer cosmological-type horizon over broad sampled sectors, with larger $c_q$ tending to push the inner horizon outward and the cosmological horizon inward; the latter showed the stronger displacement.  The same study found competing optical trends: at fixed polymer parameter, increasing $c_q$ enlarged the horizon scale and strengthened the confinement of the displayed null rays, whereas increasing the polymer parameter at fixed $c_q$ reduced the horizon and shadow radii in the sampled configurations.  It also identified parameter combinations compatible with the reported shadow-size intervals of Sgr~A* and M87* \cite{Araujo2025}.  These are source-level black-hole results, not observational constraints derived for the asymmetric shell considered here.

Let $r_{h1}$ be the positive event-horizon root of $F_1$.  On $\Mtwo$, let $r_{h2}$ and $r_c$ denote, respectively, the inner and outer positive roots that bound the selected connected interval with $F_2>0$.  The retained domains are therefore $r_1\in(r_{h1},\infty)$ and $r_2\in(r_{h2},r_c)$.  The existence, number, and ordering of these roots depend jointly on $(\bar\lambda,\bar c_q,w_q)$; labels such as ``event'' and ``cosmological'' are assigned only after the sign of the lapse and the root ordering have been checked.  In particular, no asymptotic observer, celestial screen, or emitting disk is assigned beyond $r_c$.

The cosmological horizon is not a curvature singularity and must not be treated as an arbitrary numerical cutoff.  It is the null boundary of the static chart used for $\Mtwo$: static coordinate time diverges there, while the affine radial velocity of a nonturning null ray remains finite.  The consequences are twofold.  First, the shell must lie in the same connected positive-lapse component as the selected second-side photon sphere.  Second, a ray that reaches $r_c$ has exhausted the adopted static-domain calculation, but it has not thereby been shown to escape into another asymptotic universe.  This distinction will determine the terminal ray class in Sec.~\ref{sec:ray-topology}.

\section{Static asymmetric shell and optical matching}
\label{sec:shell}

We retain $r_1\geq a$ from $\Mone$ and $a\leq r_2<r_c$ from $\Mtwo$, and identify their timelike boundaries at the common coordinate value $a$.  Because both sides use the same $H(r)$, their angular induced geometries match without an additional radial remapping.  For a temporarily dynamical embedding $a=a(\tau)$, proper-time normalization on each side gives $-F_i(a)\dot t_i^{2}+\dot a^2/F_i(a)=-1$.  The induced metric is consequently
\begin{equation}
 \dd s_\Sigma^2=-\dd\tau^2+H(a)
 \left(\dd\theta^2+\sin^2\theta\,\dd\phi^2\right).
 \label{eq:induced-metric}
\end{equation}
The physical throat radius is $R_\Sigma=\sqrt{H(a)}=\sqrt{a^2+\ell^2}$ and its area is $4\pi H(a)$; the coordinate $a$ alone is not a physical circumference radius.  A static shell is timelike only when both $F_1(a)$ and $F_2(a)$ are positive, which requires $a$ to lie above both inner horizons and below $r_c$.

Although the shell matter is not the subject of the optical calculation, the static Israel junction condition provides a useful existence check.  In the orthonormal frame comoving with the shell, the surface energy density is
\begin{equation}
 \sigma_0=-\frac{H'(a)}{8\pi H(a)}
 \left[\sqrt{F_1(a)}+\sqrt{F_2(a)}\right].
 \label{eq:static-surface-density}
\end{equation}
In geometrized units, $\sigma_0$ has dimension $\mathsf{L}^{-1}$.  For the positive-$a$ branch, $H'(a)=2a>0$, so every admissible static throat has $\sigma_0<0$.  This identifies the surface layer that supports the prescribed surgery.  Radial perturbations of the related reflection-symmetric pure-polymer and polymer--Kiselev shells have been studied for several surface equations of state \cite{Javed2022,Reboucas2026PolymerQuintessence}; in particular, the latter work found finite local stability sectors for the variable Chaplygin closure.  Here the junction quantities fix the static optical geometry, while the reflection asymmetry is carried into the null matching through the two distinct lapse values at the throat.

Optical matching is most transparent in the local orthonormal frames of static shell observers.  Their temporal basis vector is proportional to $F_i(a)^{-1/2}\partial_{t_i}$, while the azimuthal basis vector is common because the angular metric agrees.  Continuity of the locally measured photon frequency and of the momentum tangent to the shell therefore gives
\begin{equation}
 \frac{E_1}{\sqrt{F_1(a)}}=
 \frac{E_2}{\sqrt{F_2(a)}},
 \qquad
 L_1=L_2\equiv L.
 \label{eq:conserved-matching}
\end{equation}
The Killing energies $E_i$ need not agree because the two static times have different normalizations at the shell.  This is a coordinate-energy discontinuity, not a discontinuity in the frequency measured by the shell.

Writing $b_i=L/E_i$, the conserved matching becomes
\begin{equation}
 b_1=Zb_2,
 \qquad
 Z=\sqrt{\frac{F_2(a)}{F_1(a)}}.
 \label{eq:impact-matching}
\end{equation}
The factor $Z$ is real and positive precisely on the timelike static branch.  Since the observer lies at the normalized infinity of $\Mone$, $b_1$ labels the physical screen, whereas $b_2$ is an intrinsic coordinate impact used only along the second-side segment.  Thus any critical impact generated in $\Mtwo$ must be multiplied by $Z$ before it can be compared with a feature on the observer screen.

The optical topology of interest imposes one additional ordering: the throat must lie inside the selected unstable photon sphere on each side.  For the present positive branch this means $\max(r_{h1},r_{h2})<a<\min(r_{{\rm ph}1},r_{{\rm ph}2})$ together with $a<r_c$.  The upper bound is not a universal requirement for a thin-shell wormhole; it selects the configuration in which a ray can cross the shell, encounter the second-side potential barrier, and return.  In the reflection-symmetric limit $c_q\to0$, the two lapses coincide, $Z\to1$, and the two mapped critical edges merge.

\section{Null geodesics and matched critical structure}
\label{sec:null-geodesics}

Spherical symmetry allows every null ray to be placed in the equatorial plane.  With an affine parameter $\lambda_{\rm aff}$, stationarity and axial symmetry yield the constants $E_i=F_i\dot t_i$ and $L=H\dot\phi$, where a dot denotes differentiation with respect to $\lambda_{\rm aff}$.  The null constraint then reduces to
\begin{equation}
 \dot r_i^{2}=E_i^2\left[1-b_i^2\Popt_i(r_i)\right],
 \qquad
 \Popt_i(r)=\frac{F_i(r)}{H(r)}.
 \label{eq:null-radial}
\end{equation}
The dimensionless product $b_i^2\Popt_i$ controls the radial motion.  A ray is allowed where it does not exceed unity; a simple equality defines a turning point.  The common angular function means that the asymmetry enters $\Popt_i$ only through the lapse, but its effect is not a uniform vertical shift: the Kiselev term changes both the height and the location of the second-side maximum and also changes the matching factor evaluated at the throat.

An exterior circular null orbit is selected by
\begin{equation}
 F_i'(r_{{\rm ph}i})H(r_{{\rm ph}i})
 -F_i(r_{{\rm ph}i})H'(r_{{\rm ph}i})=0,
 \qquad
 b_{ci}=\sqrt{\frac{H(r_{{\rm ph}i})}{F_i(r_{{\rm ph}i})}}.
 \label{eq:photon-sphere-critical-impact}
\end{equation}
Only a root inside the retained positive-lapse component and satisfying $\Popt_i''(r_{{\rm ph}i})<0$ is used.  The second condition identifies an unstable maximum, for which the bending grows without bound as $b_i$ approaches $b_{ci}$.  Other stationary roots, if present, have a different stability or lie outside the adopted domain and cannot be inserted into the screen classification merely because they solve the first derivative equation.

To compare the two maxima on one observer screen, introduce a signed coordinate $x$ with $r_1=a+x$ for $x\geq0$ and $r_2=a-x$ for $x\leq0$.  The matched optical potential is
\begin{equation}
 \widehat{\Popt}(x)=
 \begin{cases}
  \Popt_1(a+x),&x\geq0,\\[2pt]
  Z^{-2}\Popt_2(a-x),&x\leq0.
 \end{cases}
 \label{eq:matched-potential}
\end{equation}
Equation~\eqref{eq:impact-matching} makes this potential continuous at the shell: both one-sided values equal $F_1(a)/H(a)$.  Its derivative generally has a cusp because the radial derivatives of the two lapses differ.  The observer-side maximum defines the outer critical edge $b_{\rm out}=b_{c1}$, whereas the second-side maximum maps to the inner edge $b_{\rm in}=Zb_{c2}$.  A nonempty returned-ray interval exists only when $b_{\rm in}<b_{\rm out}$.

\begin{figure*}[!htp]
\centering
\includegraphics[width=\textwidth]{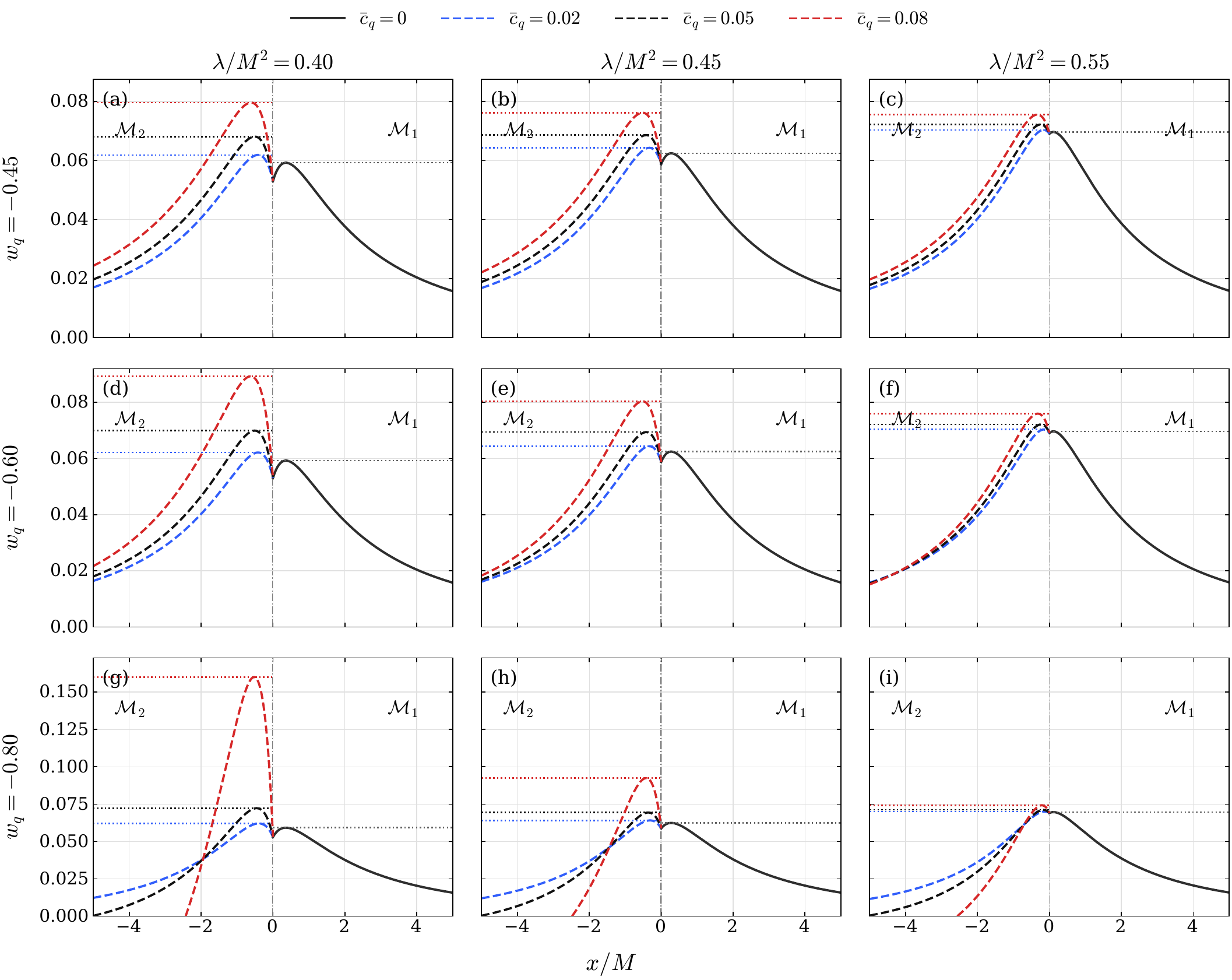}
\caption{Observer-normalized matched optical potential $M^2\widehat{\Popt}(x)$ as a function of the signed radial coordinate $x/M$, with $\Mtwo$ at $x<0$, $\Mone$ at $x>0$, and the throat at $x=0$. The exteriors have equal mass parameters, $M_1=M_2\equiv M$, and the throat is fixed at $a/M=1.8$. Rows correspond to $w_q=-0.45$, $-0.60$, and $-0.80$, while columns correspond to $\lambda/M^2=0.40$, $0.45$, and $0.55$. The solid black branch is the pure-polymer observer exterior, for which $c_q=0$, and the dashed blue, black, and red branches are the polymer--quintessence exterior for $\bar c_q=0.02$, $0.05$, and $0.08$, respectively, with $\bar c_q\equiv c_q/M^{3w_q+1}$. The vertical dash-dotted line marks the throat, and the horizontal dotted segments give the critical potential levels at the corresponding maxima. Each second-side curve is restricted to its connected positive-lapse domain and therefore ends at the cosmological horizon whenever that boundary lies inside the displayed interval.}
\label{fig:matched-potential-matrix}
\end{figure*}

Figure~\ref{fig:matched-potential-matrix} makes the reflection asymmetry explicit in the physical normalization of the observer screen. All second-side branches meet the pure-polymer branch at the throat because the factor $Z^{-2}$ in Eq.~\eqref{eq:matched-potential} implements the locally derived impact map in Eq.~\eqref{eq:impact-matching}; over the full 27-point scan, the largest numerical mismatch between the two one-sided throat values is only $1.4\times10^{-17}$. The cusp at $x=0$ is therefore a finite derivative jump produced by joining inequivalent lapses, not a discontinuity in the potential. The solid branch is unchanged down each column because varying $(\bar c_q,w_q)$ modifies only $\Mtwo$. This repeated observer-side profile is a useful control: any row dependence on the $x>0$ side would indicate that the one-sided deformation or the matching prescription had been implemented inconsistently.

At fixed $(w_q,\lambda)$, increasing $\bar c_q$ moves the second-side maximum to larger $r_2$, lowers the matching factor $Z$, and shifts the mapped critical edge $Zb_{c2}$ toward smaller observer-screen impact. The resulting separation from $b_{c1}$ grows monotonically for every row and column sampled here, even though the intrinsic second-side critical impact need not follow the same screen ordering before multiplication by $Z$. The effect can be substantial: for $w_q=-0.80$ and $\lambda/M^2=0.40$, the returned interval widens from approximately $0.095M$ at $\bar c_q=0.02$ to $1.608M$ at $\bar c_q=0.08$. The higher red barrier in panel (g) and its much lower mapped critical level are thus two representations of the same matched geometry: the former is read in the potential, whereas the latter is read on the observer screen.

The finite extent of $\Mtwo$ is equally important. The second-side curve ends where $F_2$ vanishes at the cosmological horizon, not where the numerical grid or plotted potential becomes unreliable. For the weakest large-radius deformation, $w_q=-0.45$ and $\bar c_q=0.02$, the computed outer root lies at $r_c/M\simeq7.15\times10^4$ and is far outside the displayed window. By contrast, for $w_q=-0.80$ and $\bar c_q=0.08$, it lies between $r_c/M\simeq4.23$ and $4.36$ across the three columns, so the red branches terminate visibly on the left. This comparison also exposes why $w_q$ cannot be assigned a universal monotonic effect from the matrix. At $\lambda/M^2=0.40$ and $\bar c_q=0.08$, the reflected width increases from about $0.564M$ to $0.761M$ and then $1.608M$ as $w_q$ decreases through the three rows; at $\lambda/M^2=0.55$, the corresponding sequence is approximately $0.151M$, $0.160M$, and $0.118M$. Changing the radial power of the Kiselev term modifies both the throat normalization and the neighborhood of the second photon sphere, and these contributions compete rather than producing a rigid displacement.

The columns isolate a different mechanism because $\lambda$ acts on both exteriors and on the common nonareal function $H$. Increasing $\lambda/M^2$ from $0.40$ to $0.55$ moves the observer-side photon sphere from $r_{{\rm ph}1}/M\simeq2.161$ to $1.912$, raises its potential maximum, and reduces $b_{c1}/M$ from approximately $4.108$ to $3.788$. For every sampled pair $(w_q,\bar c_q)$, the two mapped critical edges approach one another across the same column sequence, so the returned interval narrows. It nevertheless remains positive at all 27 admissible points; its smallest value, about $0.0101M$, occurs for $w_q=-0.80$, $\lambda/M^2=0.55$, and $\bar c_q=0.02$. Thus the strongest polymer correction considered here suppresses, but does not eliminate, the returned topology selected by this parameter gate.

These results delimit null turning-point structure rather than radiative visibility. A wider interval between $Zb_{c2}$ and $b_{c1}$ increases the screen domain in which returned rays are geometrically possible, but it does not determine how many disk intersections survive, how strongly they are redshifted, or whether a narrow feature is resolvable. Those questions require the orbit numbers, transfer functions, and emission laws introduced below. The numerical trends quoted here are therefore restricted to the displayed, fully admissible scan; in particular, the competing behavior with $w_q$ rules out a global monotonic extrapolation from a single row or column.

The matrix is used as a finite sensitivity survey, whereas the single configuration below is retained as a separate reference point for a quantity-by-quantity comparison of the two exteriors; it is not intended to coincide with one of the matrix panels.

The baseline fixes $M=1$, $\lambda/M^2=0.5$, $w_q=-2/3$, $c_qM=0.05$, and $a/M=1.8$.  For this value of $w_q$, the quintessence correction is simply $-c_qr$.  Positive horizon roots and stationary points were first located with sign-changing brackets and then refined with Brent's method; photon-sphere candidates were filtered by the positive-lapse, retained-domain, and instability conditions.  The resulting comparison is shown in Table~\ref{tab:critical-baseline}.  Rows are the two exteriors and columns are the physically comparable quantities, so the effect of activating quintessence is visible directly.

\begin{table}[!htp]
\caption{Critical structure of the baseline configuration.  The last column maps each intrinsic critical impact to the observer screen; hence the $\Mtwo$ entry contains $Zb_{c2}$ rather than $b_{c2}$.  A dash denotes a quantity not used for that exterior.}
\label{tab:critical-baseline}
\begin{ruledtabular}
\begin{tabular}{lcccccc}
Exterior & $r_h/M$ & $r_c/M$ & $r_{\rm ph}/M$ & $b_c/M$ & $F_i(a)$ & screen edge$/M$\\
\hline
$\Mone$ (polymer) & $1.2166174$ & --- & $1.9949907$ & $3.8948975$ & $0.2480407$ & $3.8948975$\\
$\Mtwo$ (polymer--quintessence) & $1.3416505$ & $17.8760320$ & $2.1131655$ & $4.7289722$ & $0.1580407$ & $3.7747638$\\
\end{tabular}
\end{ruledtabular}
\end{table}

The table exhibits three distinct quintessence effects.  First, the inner horizon and photon sphere move outward relative to their pure-polymer counterparts, consistently with the source trends for increasing $c_q$ in the relevant sampled sector \cite{Araujo2025}.  Second, the cosmological horizon makes the second static region finite.  Third, the lower throat lapse gives $Z=0.7982208$, so the larger intrinsic value of $b_{c2}$ maps to the smaller screen radius $b_{\rm in}/M=3.7747638$.  Comparing $b_{c2}$ directly with $b_{c1}$ would therefore reverse the physical ordering seen by the observer.

The screen interval available for return has width $\Delta b_{\rm ref}/M=0.1201338$, only about $3.08\%$ of $b_{c1}$.  This is already a validated geometric statement: the baseline contains two nearby critical mechanisms, one generated locally by the observer-side photon sphere and one generated contralaterally and transported through the shell.  It is not yet a statement about two visible rings.  A transfer branch must also reach the emitting part of the disk, its screen support must be resolved, and its redshifted intensity must survive the adopted source model.

The baseline ordering is $r_{h2}<a<r_{{\rm ph}1}<r_{{\rm ph}2}<r_c$, while $r_{h1}<a$ on the observer side.  It therefore satisfies both the timelike-shell condition and the returned-ray ordering.  The root and matching residuals have been independently reproduced by the project geometry benchmark.  In the limit $c_q\to0$, $Z$, the two photon spheres, and the two intrinsic critical impacts become equal, so $\Delta b_{\rm ref}$ collapses to zero.  The finite returned interval is thus caused by reflection asymmetry, not by the mere presence of a shell.

\section{Ray topology and nonareal angular propagation}
\label{sec:ray-topology}

The two mapped critical edges divide the observer screen into three mutually exclusive noncritical ray classes.  Rays with $b_1>b_{c1}$ turn in $\Mone$ and return without sampling the shell.  Rays with $Zb_{c2}<b_1<b_{c1}$ cross into $\Mtwo$, turn outside its photon sphere, recross the shell, and reach the observer.  Rays with $0<b_1<Zb_{c2}$ cross the shell but possess no second-side turning point inside the static patch; they propagate to $r_c$.  Exactly critical rays asymptotically approach the corresponding unstable photon orbit and form the limiting boundaries between these classes.

The distinction is topological rather than merely numerical.  The first class depends only on $F_1$ and is therefore unchanged when $(c_q,w_q)$ vary at fixed $\Mone$.  The returned class depends on $\Mtwo$ twice: its screen support is set by $Zb_{c2}$, and its accumulated angle depends on the entire second-side potential between the throat and the turning point.  The terminal class also probes $\Mtwo$, but it does not return to generate an outgoing disk-intersection sequence.  Because no source is placed on that side, reaching $r_c$ adds no contralateral emission.

The angular propagation equation follows directly from Eq.~\eqref{eq:null-radial},
\begin{equation}
 \left|\frac{\dd\phi}{\dd r_i}\right|
 =\frac{b_i}{H(r_i)\sqrt{1-b_i^2F_i(r_i)/H(r_i)}}.
 \label{eq:angular-quadrature}
\end{equation}
The explicit factor $H(r)=r^2+\ell^2$ is essential: replacing it by $r^2$ would alter not only the centrifugal barrier but also the angular weight assigned to every radial interval.  Rather than repeating the same integrand for each ray class, define the positive angular advance along any allowed segment by
\begin{equation}
 \begin{aligned}
 \mathcal{A}_i(r_-,r_+;b_i)
 &\equiv\int_{r_-}^{r_+}
 \frac{b_i\,\dd r}{H(r)\sqrt{1-b_i^2F_i(r)/H(r)}},\\
 \Phi_1(b_1)&\equiv\mathcal{A}_1(a,\infty;b_1).
 \end{aligned}
 \label{eq:angular-segment}
\end{equation}
Here $r_-<r_+$ are endpoints in the retained static domain and the radicand is nonnegative along the open segment.  The functional $\mathcal{A}_i$ is dimensionless and records only the amount of azimuth accumulated; the direction of radial motion is supplied separately by the ray topology.  The second line is the one-way observer-to-throat contribution shared by every subcritical ray.  It depends only on $F_1$, $H$, and $b_1$, so changing quintessence on $\Mtwo$ cannot modify the incoming trajectory before it reaches the shell.

What happens after crossing is determined by the mapped inner critical edge.  With $b_2=b_1/Z$, the second-side contribution can be written without introducing a new integrand:
\begin{equation}
 \Phi_2(b_1)=
 \begin{cases}
  \Phi_2^{c}\equiv\mathcal{A}_2(a,r_c;b_1/Z),
  &0<b_1<Zb_{c2},\\[2pt]
  \Phi_2^{\rm rt}\equiv2\mathcal{A}_2(a,r_{t2};b_1/Z),
  &Zb_{c2}<b_1<b_{c1}.
 \end{cases}
 \label{eq:second-side-angular-branches}
\end{equation}
In the first branch, the photon has no turning point in the retained part of $\Mtwo$ and its angular evolution ends at the cosmological horizon.  Because $F_2\to0$ at a simple $r_c$, the square root in Eq.~\eqref{eq:angular-segment} tends to unity there: the angular advance is finite even though the static coordinate time diverges.  The horizon is therefore a regular endpoint for this angular bookkeeping, but not evidence of escape to a second asymptotic region or of any particular continuation beyond the static chart.

The second branch contains a genuine round trip.  Its outer turning point $r_{t2}$ is the root of $(b_1/Z)^2F_2(r_{t2})/H(r_{t2})=1$ between $r_{{\rm ph}2}$ and $r_c$, and the factor of two counts the ingoing and outgoing traversals of the same radial segment.  As $b_1\to Zb_{c2}^{+}$, the turning point approaches the unstable second-side photon sphere and $\Phi_2^{\rm rt}$ grows logarithmically.  This large accumulation reflects the prolonged angular dwell near the unstable orbit and compresses higher-order returned trajectories near the inner screen edge; it is not a singularity of the throat or a numerical duplication of one segment.

The remaining topology never reaches the shell.  For $b_1>b_{c1}$, the observer-side turning point $r_{t1}>r_{{\rm ph}1}$ satisfies $b_1^2F_1(r_{t1})/H(r_{t1})=1$, and the complete scattering advance is
\begin{equation}
 \Phi_1^{\rm sc}(b_1)=2\mathcal{A}_1(r_{t1},\infty;b_1).
 \label{eq:scattering-angle}
\end{equation}
Here the factor of two again has a topological meaning, now accounting for approach to and departure from the observer-side turning point.  The divergence as $b_1\to b_{c1}^{+}$ and the growth of $\Phi_1$ as $b_1\to b_{c1}^{-}$ are the two one-sided manifestations of the same unstable orbit in $\Mone$, but they belong to scattered and shell-crossing rays, respectively.  By contrast, the divergence at $Zb_{c2}$ is created only after matching into $\Mtwo$.  This separation also clarifies the parameter dependence: $(\bar c_q,w_q)$ change the second-side branches through both $F_2$ and $Z$, while leaving the observer-side segment functional unchanged; $\lambda$, however, deforms $F_1$, $F_2$, and the common nonareal function $H$, and can therefore shift all three angular mechanisms.

\begin{figure*}[!htp]
\centering
\includegraphics[width=1\textwidth]{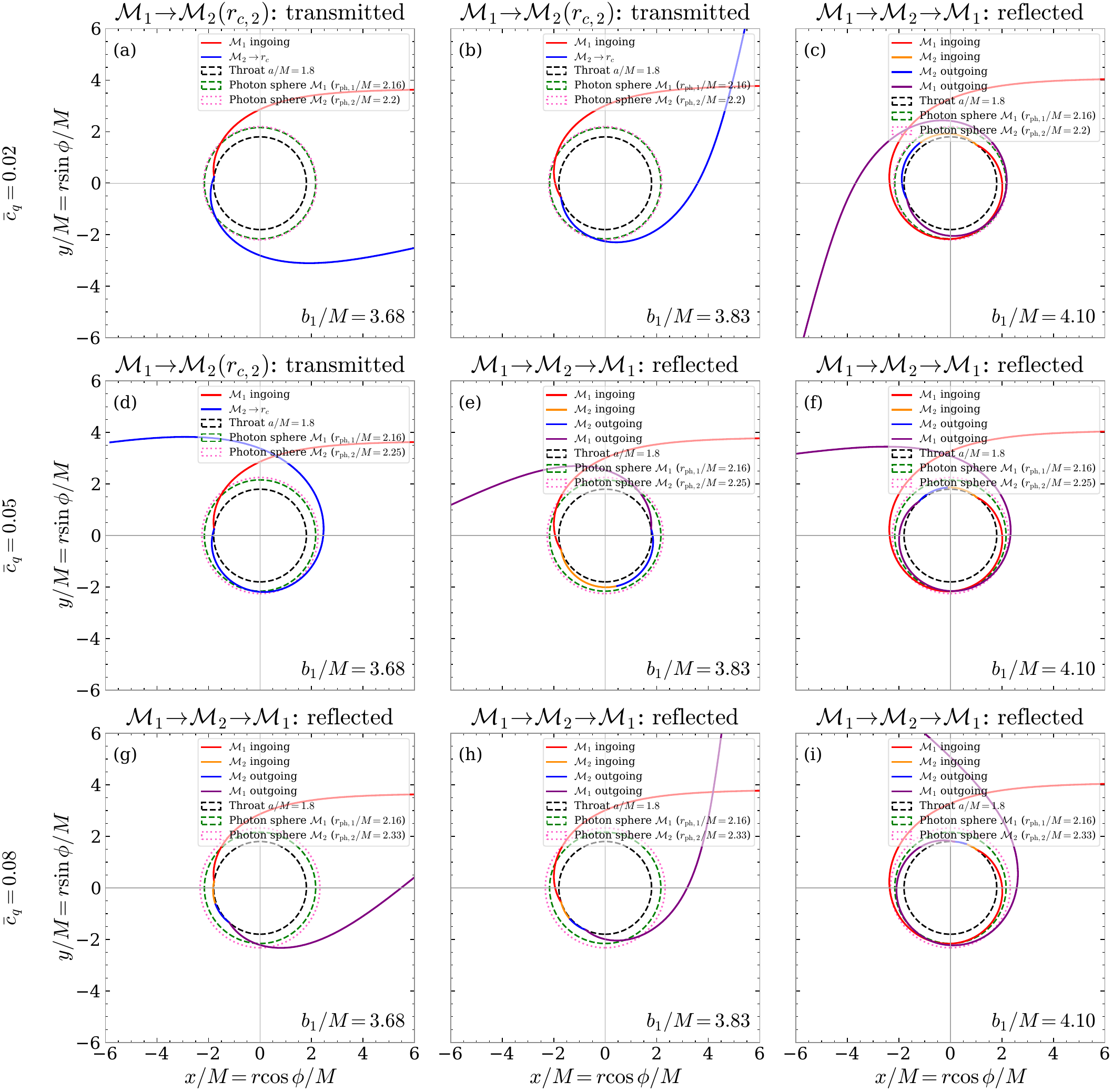}
\caption{Representative backward-traced null trajectories for $M_1=M_2\equiv M$, $\lambda/M^2=0.40$, $a/M=1.8$, and $w_q=-0.80$. Rows correspond to $\bar c_q=0.02$, $0.05$, and $0.08$, while columns fix the observer-screen impact at $b_1/M=3.68$, $3.83$, and $4.10$. Panel titles identify whether a crossed ray is transmitted to the second-side cosmological horizon or reflected back through the throat. The red segment is ingoing in $\Mone$; orange and blue denote the ingoing and outgoing portions in $\Mtwo$, respectively; purple is the final outgoing segment in $\Mone$. The dashed black circle marks the throat, while the dashed green and dotted magenta circles mark the photon spheres of $\Mone$ and $\Mtwo$. The Cartesian window $-6\leq x/M,y/M\leq6$ is a display crop: every transmitted branch is integrated to the one-sided limit at $r_{c,2}$ even when it first leaves the frame.}
\label{fig:null-trajectories-matrix}
\end{figure*}

Figure~\ref{fig:null-trajectories-matrix} realizes the screen partition derived from the two matched critical edges. For the common observer exterior at $\lambda/M^2=0.40$, the outer edge is $b_{\rm out}/M=b_{c1}/M\simeq4.107985$, so all three displayed impacts remain below the observer-side scattering threshold and reach the throat. The inner edge, by contrast, decreases from $b_{\rm in}/M\simeq4.012743$ to $3.719053$ and $2.499868$ along the three rows. Consequently, the first row contains two transmitted rays followed by one reflected ray, the middle row contains one transmitted and two reflected rays, and the bottom row is entirely reflected. The matrix therefore samples the topology changes caused by the second exterior while keeping the observer-side access to the shell fixed.

The repeated red segment within each column is a nontrivial control of this separation. Before crossing the shell, the ray depends only on $F_1$, $H$, and $b_1$, none of which changes between rows. After matching through Eq.~\eqref{eq:impact-matching}, increasing $\bar c_q$ lowers $Z$ from approximately $0.879$ to $0.658$ and then $0.303$, while the second-side photon sphere moves outward from $r_{{\rm ph}2}/M\simeq2.195$ to $2.254$ and $2.325$. The pronounced downward shift of $Zb_{c2}$ is therefore not determined by the modest displacement of the magenta circle alone; it combines the altered second-side barrier with the increasingly different time normalization at the throat. This is the trajectory-level manifestation of the widening returned interval found in Fig.~\ref{fig:matched-potential-matrix}. Each row represents a distinct static geometry, rather than the temporal evolution of one photon or one surrounding medium.

The transmitted panels also make the finite causal domain of $\Mtwo$ operational. A transmitted ray has no second-side turning point and is followed only to the cosmological horizon, whose radius decreases from $r_{c,2}/M\simeq14.830$ to $6.846$ and $4.229$ as $\bar c_q$ increases. It is not an escape ray to a second asymptotically flat infinity. The terminal branch in Eq.~\eqref{eq:second-side-angular-branches} has a finite one-sided angular limit at that boundary, and the numerical integrations reach the selected horizon with a maximum terminal-radius residual of about $7.1\times10^{-15}M$; branches that leave the Cartesian crop first are therefore not truncated physical solutions. The comparison between panels (d) and (g) is particularly instructive: bringing the horizon inward does not itself make a ray terminal. Panel (d) remains transmitted because $b_1<b_{\rm in}$, whereas panel (g) is reflected despite the still smaller static patch because the matched barrier now supplies a turning point before the horizon.

The column sequence probes complementary portions of that screen partition. At $b_1/M=3.68$, reflection occurs only for the strongest displayed quintessence amplitude; at $3.83$, it already occurs in the two lower rows. The last column has a different critical role. Its value $b_1/M=4.10$ lies only about $0.007985M$ below $b_{c1}$ and is above all three mapped inner edges, so every ray crosses, turns in $\Mtwo$, and returns while also accumulating a large observer-side deflection near the $\Mone$ photon sphere. This is a deliberately resolved near-critical sample, not an orbit placed on the unstable circular null trajectory: the fine/coarse integrations agree in endpoint azimuth to better than $6.8\times10^{-10}$, with no detected discontinuity at either shell crossing.

No scattered column is shown because scattering begins only for $b_1/M>b_{c1}/M\simeq4.107985$. Such a ray turns entirely in $\Mone$, follows Eq.~\eqref{eq:scattering-angle}, and never probes the shell, the cosmological horizon, or the quintessence potential. Since the one-sided deformation leaves $\Mone$ unchanged, the same scattered curve would be repeated in all three rows and would add no information about reflection asymmetry. The choice $b_1/M=4.10$ approaches this boundary from below and thus retains interaction with both exteriors; increasing it by slightly more than $0.008M$ would instead convert the entire column into the row-independent scattered class. The absence of mirrored or repeated observer-side trajectories is therefore intentional, not evidence that those geodesics are absent.

Under the unilateral source prescription, the distinction between transmitted and reflected rays has a direct but limited radiative consequence. A transmitted ray terminates at $r_{c,2}$ and cannot return to create a new intersection with the disk in $\Mone$, whereas a reflected ray recrosses the shell and can support additional transfer branches. The trajectory plot establishes only the existence and geometry of those paths. Their disk-intersection order, screen width, redshifted weight, and eventual resolvability must still be obtained from the transfer and intensity calculations; moreover, the specific topology pattern above is conditional on the displayed slice $(\lambda/M^2,w_q)=(0.40,-0.80)$ and should not be promoted to a global monotonic statement.

\section{Orbit numbers and transfer functions}
\label{sec:transfers}

Orbit numbers provide a compact measure of accumulated azimuth, not a brightness assignment.  For a path confined to $\Mone$, the relevant angular accumulation is the one-sided ingoing angle below the observer-side critical impact and the complete scattering angle above it.  Accordingly,
\begin{equation}
 n_1(b_1)=
 \begin{cases}
 \Phi_1(b_1)/(2\pi),&0<b_1<b_{c1},\\[3pt]
 \Phi_1^{\rm sc}(b_1)/(2\pi),&b_1>b_{c1}.
 \end{cases}
 \label{eq:orbit-number-side1}
\end{equation}
The critical value $b_1=b_{c1}$ is excluded because both branches diverge logarithmically there.
For returned rays, path ordering yields the two additional counts used in asymmetric thin-shell imaging \cite{Peng2021},
\begin{equation}
 n_2=\frac{\Phi_1+\Phi_2^{\rm rt}}{2\pi},
 \qquad
 n_3=\frac{2\Phi_1+\Phi_2^{\rm rt}}{2\pi}.
 \label{eq:returned-orbit-numbers}
\end{equation}
The intervals $n<3/4$, $3/4<n<5/4$, and $n>5/4$ conventionally distinguish direct, lensing-ring, and photon-ring angular sectors \cite{Gralla2019}.  These thresholds classify how much azimuth a trajectory accumulates.  They do not determine whether a corresponding branch intersects the disk, carries nonzero emission, or is observable after finite resolution.

\begin{figure}[!htp]
\centering
\includegraphics[width=0.5\columnwidth]{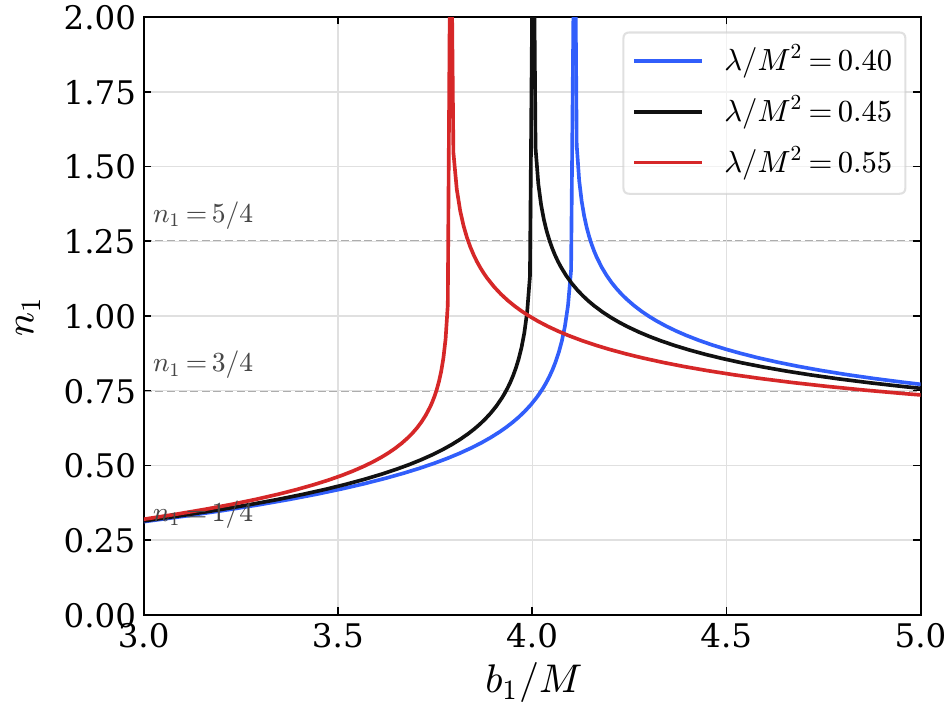}
\caption{Observer-side orbit number $n_1$ as a function of the physical screen impact parameter $b_1/M$ for $\lambda/M^2=0.40$, $0.45$, and $0.55$, with $M_1=M_2\equiv M$ and $a/M=1.8$. The subcritical branch gives the one-way angular advance from the observer to the throat, whereas the supercritical branch gives the complete scattering advance in $\Mone$. The horizontal lines delimit the direct, lensing-ring, and photon-ring angular sectors. The apparent termination at $n_1=2$ is the plotting boundary: both branches continue to larger orbit number as $b_1$ approaches the corresponding $b_{c1}$.}
\label{fig:orbit-number-n1-lambda}
\end{figure}

Figure~\ref{fig:orbit-number-n1-lambda} isolates the optical response generated entirely in the observer exterior. For $b_1<b_{c1}$, the backward-traced photon reaches the throat and $n_1$ measures the one-way angular accumulation defined by Eq.~\eqref{eq:angular-segment}. For $b_1>b_{c1}$, a radial turning point forms in $\Mone$, and the plotted quantity becomes the complete scattering advance of Eq.~\eqref{eq:scattering-angle}. The two sides of each vertical rise consequently describe different topologies rather than a single curve continued through the critical impact. On either side, the turning structure approaches the unstable photon sphere selected by Eq.~\eqref{eq:photon-sphere-critical-impact}, producing the familiar near-critical growth. The upper cutoff at $n_1=2$ is therefore purely graphical; the exactly critical trajectory approaches the circular null orbit asymptotically and does not complete an infinite number of revolutions in finite propagation.

The polymer parameter relocates this common critical structure over a sizeable portion of the observer screen. Across the displayed sequence $\lambda/M^2=(0.40,0.45,0.55)$, the independently computed photon-sphere radii decrease as $r_{{\rm ph}1}/M=(2.1607,2.0779,1.9116)$, while the associated critical impacts decrease as $b_{c1}/M=(4.1080,4.0013,3.7883)$. The full pair of branches therefore moves toward smaller $b_1/M$ as $\lambda$ increases. This displacement is not attributable to a change in an areal centrifugal term alone: $\lambda$ modifies both the observer-side lapse in Eq.~\eqref{eq:pure-polymer-lapse} and the nonareal angular function in Eq.~\eqref{eq:angular-function}. Their ratio fixes the photon barrier, while the entire radial profiles of both functions enter the accumulated angle. The separation among the curves remains comparatively modest away from the barrier, but becomes strongly magnified near $b_{c1}$ because a small screen displacement then corresponds to a long residence near the unstable orbit.

This motion also shifts the intersections with the $n_1=3/4$ and $5/4$ thresholds, thereby relocating the screen intervals assigned to lensing-ring- and photon-ring-order angular accumulation before any radiative model is specified. The plot shows this reorganization only over $3\leq b_1/M\leq5$ and clips the divergent portions at $n_1=2$; neither boundary represents the end of a physical branch. Conversely, the finite scan establishes the ordering and critical displacement within the sampled polymer range, but does not by itself justify extrapolating their monotonicity beyond it. The inclusion of $\lambda/M^2=0.40$ is especially useful because it extends the observer-side comparison toward weaker polymer deformation while preserving the same two-branch topology.

No separate scan in $\bar c_q$ or $w_q$ is shown because it would reproduce these curves exactly. Both quintessence parameters occur only in $F_2$, whereas the quadratures defining the two $n_1$ branches involve $F_1$ and $H$ alone: they contain neither the matching factor $Z$ nor a path segment in $\Mtwo$. The insensitivity of $n_1$ to quintessence is thus an analytic consequence of the one-sided construction, not a numerical degeneracy or an effect hidden by the plotting resolution. This single panel serves accordingly as an observer-side control. Dependence on $(\bar c_q,w_q)$ can enter the terminal and returned families after throat crossing, but cannot alter the ordinary $n_1$ hierarchy while $\Mone$ and the throat location are held fixed.

The horizontal thresholds organize the angular bookkeeping of the ray family. A trajectory with $n_1>5/4$ has accumulated a photon-ring-order angle, while its radiative expression is selected by the intersection condition in Eq.~\eqref{eq:transfer-functions}, the radial support of the emissivity, the transfer slope, and the redshift weighting. Figure~\ref{fig:orbit-number-n1-lambda} therefore fixes how the polymer observer geometry organizes the ordinary critical hierarchy, and the subsequent transfer and intensity analyses map that hierarchy into source-weighted image features.

\begin{figure*}[!htp]
\centering
\includegraphics[width=\textwidth]{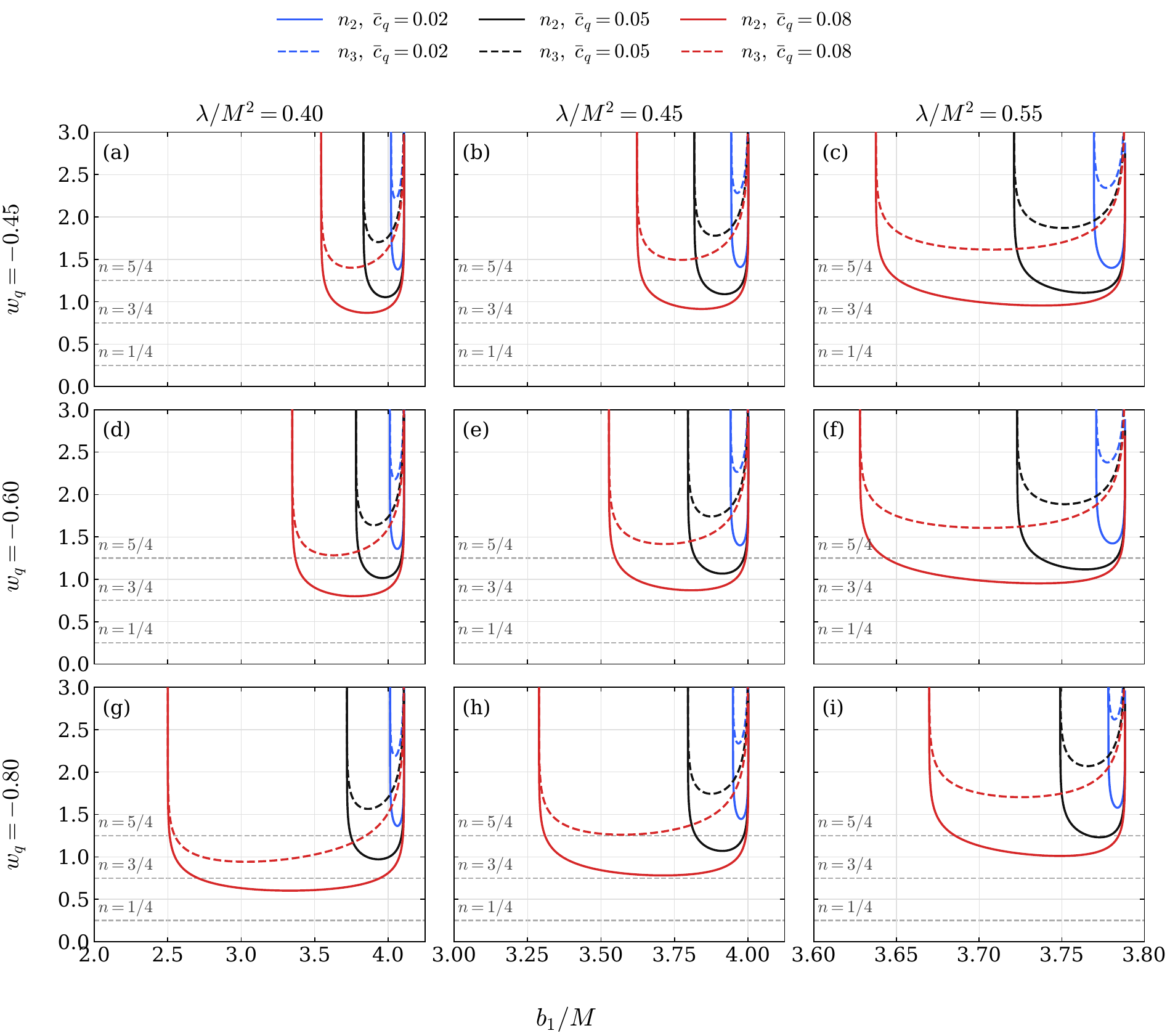}
\caption{Returned-ray orbit numbers $n_2$ (solid) and $n_3$ (dashed) as functions of the physical observer-screen impact $b_1/M$, with $M_1=M_2\equiv M$ and $a/M=1.8$. Rows correspond to $w_q=-0.45$, $-0.60$, and $-0.80$, columns to $\lambda/M^2=0.40$, $0.45$, and $0.55$, and colors to $\bar c_q=0.02$, $0.05$, and $0.08$, where $\bar c_q\equiv c_q/M^{3w_q+1}$. The column windows are, respectively, $2\leq b_1/M\leq4.25$, $3\leq b_1/M\leq4.125$, and $3.6\leq b_1/M\leq3.8$. Each curve is restricted to its own returned interval $Zb_{c2}<b_1<b_{c1}$; its growth at the left and right endpoints is associated with the side-2 and side-1 unstable photon spheres, respectively. The horizontal lines delimit the conventional angular sectors, and the curves continue above the common displayed ceiling $n_{2,3}=3$ near both critical edges.}
\label{fig:orbit-numbers-n2n3-matrix}
\end{figure*}

The two-ended profiles in Fig.~\ref{fig:orbit-numbers-n2n3-matrix} are the integral signature of a returned photon sampling both exterior barriers. As $b_1$ approaches the mapped inner edge $Zb_{c2}$ from above, the turning point in $\Mtwo$ approaches its unstable photon sphere and the round-trip branch in Eq.~\eqref{eq:second-side-angular-branches} grows without bound. At the opposite edge, $b_1\to b_{c1}^{-}$, the observer-side segments dwell increasingly close to the unstable orbit in $\Mone$. The finite minimum between the two rises identifies the least-wound member of that returned family; it is not a stable light ring or a dynamically preferred orbit. The dashed curve remains above its solid partner because Eq.~\eqref{eq:returned-orbit-numbers} assigns $n_3$ one additional positive observer-side segment after the same contralateral round trip. The clipping at $n_{2,3}=3$ therefore hides only the continuation of the critical growth, not a physical termination of either family.

The most systematic asymmetric trend is controlled by the quintessence amplitude. At fixed $(w_q,\lambda)$, increasing $\bar c_q$ moves the mapped second-side edge to smaller $b_1/M$ without changing the observer edge, broadening the returned interval and lowering the minima of both orbit-number families in all nine panels. The numerically determined sector changes are correspondingly sharp. At $\bar c_q=0.02$, every $n_2$ minimum remains above $5/4$; at $\bar c_q=0.05$, all nine $n_2$ minima lie between $3/4$ and $5/4$. For $\bar c_q=0.08$, the $n_2$ minimum remains in that lensing-ring-order sector except in panel (g), where it falls to approximately $0.602$. The $n_3$ minima stay above $5/4$ throughout the matrix except in the same panel, where the additional observer-side segment raises the minimum only to approximately $0.942$. Stronger quintessence therefore does not simply increase the angular advance. Through Eq.~\eqref{eq:quintessence-lapse}, it changes $F_2(a)$, the matching factor $Z$, the intrinsic second-side critical impact, and the full round-trip integrand; over this scan, the dominant combined effect is to admit a broader set of less-wound returned trajectories.

The column comparison exposes a competing polymer effect. As $\lambda/M^2$ increases from $0.40$ to $0.45$ and $0.55$, the observer edge moves from $b_{c1}/M\simeq4.1080$ to $4.0013$ and $3.7883$. The mapped inner edge also moves, but the calculated returned width contracts for every sampled $(w_q,\bar c_q)$. At $w_q=-0.60$ and $\bar c_q=0.08$, for example, it decreases from approximately $0.761M$ to $0.474M$ and $0.160M$. The orbit-number minima are generally lifted as this interval narrows, although the weak-amplitude top row is not strictly monotonic, so the effect is not a rigid horizontal compression. The full scan contains returned widths from approximately $0.0101M$ to $1.608M$; because the three columns deliberately use different horizontal windows, these quantitative values, rather than the apparent span on the page, provide the meaningful width comparison. The contraction reflects the simultaneous deformation of both lapses and of the nonareal function $H$ by $\lambda$, in contrast with the one-sided action of $\bar c_q$.

Varying $w_q$ produces an additional reversal because it changes the radial power of the Kiselev contribution rather than only its amplitude. For $\lambda/M^2=0.40$ and $\bar c_q=0.08$, the returned widths increase from about $0.564M$ to $0.761M$ and $1.608M$ as $w_q$ decreases through the three rows. At $\lambda/M^2=0.55$, the corresponding sequence is instead approximately $0.151M$, $0.160M$, and $0.118M$: the most negative displayed value now narrows the interval. Equation~\eqref{eq:quintessence-lapse} accounts for why no single ordering is expected. Changing its exponent modifies the lapse at the throat, and hence $Z$, differently from the lapse near the second photon sphere, which controls $b_{c2}$ and the round-trip accumulation. The row dependence is therefore a matched, scale-dependent competition, consistent with the critical-edge reversals already identified in Fig.~\ref{fig:matched-potential-matrix}; it cannot be reduced to the statement that a more negative $w_q$ always strengthens or weakens bending.

The sector crossings have a direct but limited implication for prospective image order. In the broadest returned window, panel (g), the least-wound $n_2$ trajectory has direct-order angular accumulation while its $n_3$ partner has lensing-ring-order accumulation; by contrast, the weak-amplitude curves remain of photon-ring order even at their minima. These are distinct static geometries selected by the parameter grid, not stages in the evolution of one wormhole. Moreover, an orbit-number sector is not itself an image: a returned path contributes only if its ordered outgoing segment reaches a target angle in Eq.~\eqref{eq:intersection-targets} at a retained emitting radius. Its eventual screen width and intensity then depend on the transfer slope, source support, and redshift. Within the sampled domain, the matrix therefore establishes that quintessence can open lower-winding contralateral channels, while $\lambda$ and $w_q$ regulate their accessibility through competing matching and propagation effects; it does not by itself establish a bright or resolvable ring.

For a face-on observer and a geometrically thin equatorial disk, the target azimuths of successive intersections are
\begin{equation}
 \phi^{(m)}=\frac{(2m-1)\pi}{2},
 \qquad m=1,2,3,\ldots.
 \label{eq:intersection-targets}
\end{equation}
The $m$th transfer function is the emitting radius $r_m(b_1)$ that solves
\begin{equation}
 \Phi_{\rm path}\!\left(r_m;b_1\right)=\phi^{(m)},
 \qquad r_m(b_1)\geq a.
 \label{eq:transfer-functions}
\end{equation}
Here $\Phi_{\rm path}$ is accumulated along the ordered ingoing, crossed, reflected, and outgoing segments appropriate to the ray class.  The restriction $r_m\geq a$ is physical: the shell excises the seed geometry and any formal disk support below it.  A branch can be born when the accumulated angle first reaches a target at the throat, can disappear when its intersection falls below an emission cutoff, and can terminate at a critical edge as its source radius or slope approaches a limiting value.

Implicit differentiation makes the geometric magnification information explicit,
\begin{equation}
 \frac{\dd r_m}{\dd b_1}
 =-\frac{\partial_{b_1}\Phi_{\rm path}}
 {\partial_{r_m}\Phi_{\rm path}}.
 \label{eq:transfer-slope}
\end{equation}
A nearly vertical transfer branch maps a broad interval of emitting radii into a narrow interval on the screen and is therefore geometrically demagnified.  Conversely, a small slope can spread a compact source range over a wider screen band.  Neither behavior fixes the received intensity by itself, because the source support and gravitational redshift still weight each intersection.

\begin{figure*}[!htp]
\centering
\includegraphics[width=\textwidth]{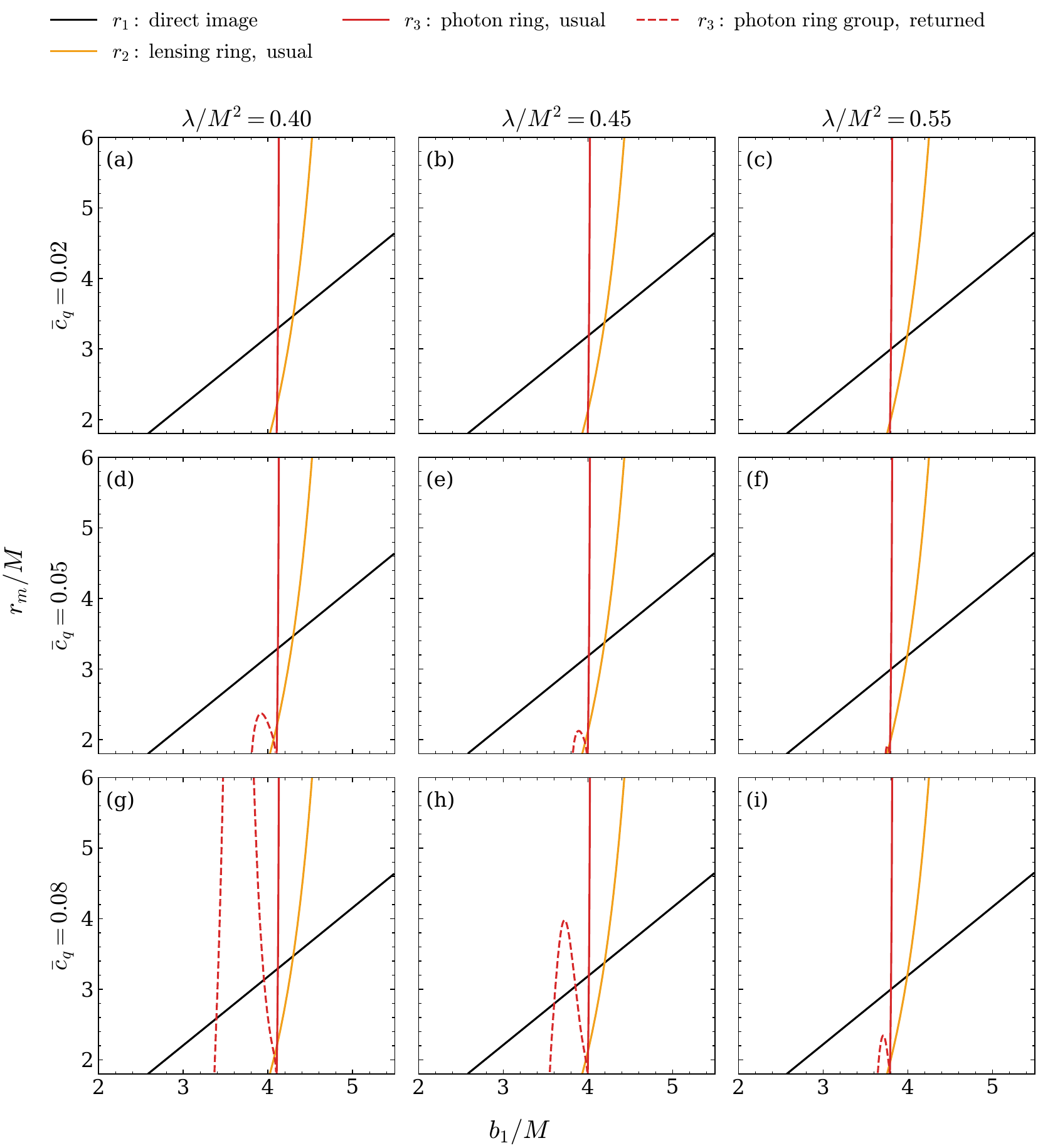}
\caption{Disk transfer functions $r_m(b_1)$ retained through third order for $M_1=M_2\equiv M$, $a/M=1.8$, and fixed $w_q=-0.60$. Rows correspond to $\bar c_q=0.02$, $0.05$, and $0.08$, with $\bar c_q\equiv c_q/M^{3w_q+1}$, while columns correspond to $\lambda/M^2=0.40$, $0.45$, and $0.55$. The horizontal coordinate $b_1/M$ is the physical impact on the observer screen, and $r_m/M$ gives the radius of the corresponding intersection with the thin disk in $\Mone$. Solid black, orange, and red curves denote the usual first-, second-, and third-order branches, respectively, and the dashed red curve denotes the returned third-order branch. No returned second-order solution exists in this parameter matrix. The legend lists only the families present within the declared truncation $m\leq3$. The lower frame boundary is the throat, so only $r_m\geq a$ is retained; the common windows $2\leq b_1/M\leq5.5$ and $1.8\leq r_m/M\leq6$ are display choices, and branches reaching the upper frame continue in the fully integrated data.}
\label{fig:transfer-functions-matrix}
\end{figure*}

Figure~\ref{fig:transfer-functions-matrix} resolves the first three ordered disk intersections selected by Eq.~\eqref{eq:transfer-functions}. The broad black branch is the direct image, whereas the usual second- and third-order branches become progressively concentrated around the observer-side critical impact. This hierarchy follows from the increasing target angles in Eq.~\eqref{eq:intersection-targets}: a higher-order intersection requires a ray that spends longer near the unstable observer-side photon orbit. The nearly vertical orange and red segments therefore have large $|\dd r_m/\dd b_1|$ and are geometrically demagnified according to Eq.~\eqref{eq:transfer-slope}. Their narrow screen support is a physical critical compression, rather than a gap produced by the sampling, but it is not by itself evidence for a bright or resolvable ring.

The row-wise coincidence of the three solid families provides an exact one-sided control. These usual intersections are encountered before the backward-traced ray can acquire any dependence on propagation through $\Mtwo$. Hence changing $\bar c_q$ at fixed $\lambda$ leaves them invariant because $F_1$, $H$, $a$, the observer, and the disk are unchanged. By contrast, increasing $\lambda/M^2$ from $0.40$ to $0.55$ deforms both the observer-side lapse and the nonareal angular function and moves the usual critical hierarchy toward smaller $b_1/M$, consistently with Fig.~\ref{fig:orbit-number-n1-lambda}. The superposed solid curves down a column are consequently identical solutions for different second exteriors, not distinct curves hidden by the resolution.

The dashed red family isolates the third-order intersection created after a ray crosses the shell, turns in $\Mtwo$, and returns to the emitting exterior. No returned second-order branch occurs in any panel. This absence follows from the ordered angular bookkeeping: throughout the sampled matrix, the smallest accumulated angle at the beginning of the final $\Mone$ segment is already about $5.03$ rad, above the second target $3\pi/2$. That target is therefore crossed before the ray can meet the observer-side disk on its return. The later third target can still be reached after recrossing, but only when it lies within the angular interval swept by the outgoing segment. Accordingly, the returned $r_3$ branch is absent for $\bar c_q=0.02$, appears close to the throat for $\bar c_q=0.05$, and gains substantially larger radial and screen support for $\bar c_q=0.08$. Its absence in the first row means that this particular disk intersection does not exist there; it does not imply the absence of returned null trajectories.

Because the transfer matrix now uses the same $w_q=-0.60$ slice as the middle row of Fig.~\ref{fig:orbit-numbers-n2n3-matrix}, their parameter dependence can be compared point by point. For $\bar c_q=0.02$, the returned critical width decreases from approximately $0.0968M$ to $0.0604M$ and $0.0174M$ as $\lambda/M^2$ increases across the columns; for $\bar c_q=0.08$, the corresponding sequence is about $0.761M$, $0.474M$, and $0.160M$. Increasing the quintessence amplitude thus broadens the set of screen impacts that return from $\Mtwo$, while increasing the polymer parameter contracts it over this fixed-$w_q$ scan. The discrete transfer consequence is visible in the dashed branch: at $\bar c_q=0.08$ its radial extent exceeds the upper frame in panel (g), remains below $r_m/M\simeq4$ in panel (h), and is confined below approximately $2.35$ in panel (i). This nonrigid response reflects the competition between the $\lambda$ dependence of both exteriors and $H$ and the one-sided modification of $F_2$, $Z$, and the round-trip angle by $\bar c_q$.

The cosmological horizon determines which crossed rays can contribute to these returned branches even though it is neither an emitter nor a plotted transfer radius. Rays below the mapped second-side critical edge reach that horizon within the adopted static chart and do not recross the throat, so they cannot produce an outgoing intersection with the disk in $\Mone$. The dashed solutions instead turn inside the finite static patch and return before reaching the horizon. The lower frame boundary $r_m/M=1.8$ is therefore physical because the shell excises smaller source radii, whereas the upper boundary $r_m/M=6$ is purely graphical. The full integration contains usual second-order radii above $600M$, usual third-order radii up to tens of $M$, and a returned third-order radius up to about $17.6M$ in the sampled matrix; curves cut by the top frame continue rather than terminate there.

The figure is intentionally restricted to $m\leq3$ and should not be read as the complete image hierarchy. Independent angular scans find usual fourth- and fifth-order intersections near the observer-side critical edge and returned fourth- and sixth-order intersections in all nine configurations, with additional orders in part of the matrix. These omitted solutions are the expected continuation of critical winding, not missing branches within the declared figure scope. Conversely, the displayed families have continuous numerical support, remain above the throat, and agree with independent direct branch solutions to within $7.1\times10^{-6}M$ at the tested points. The matrix therefore establishes the geometry and parameter dependence of the first three transfer orders only. Whether any of them contributes appreciable received intensity still depends on the emission support and on the redshift weighting in Eq.~\eqref{eq:observed-intensity}.

\section{Observer-side circular motion, emission, and optical observables}
\label{sec:emission}

The observer and the emitting disk are confined to $\Mone$.  This unilateral illumination is essential to the controlled comparison: varying $(c_q,w_q)$ modifies only the returned propagation and the terminal ray fate, while the observer-side lapse, ordinary transfer branches, and source profiles remain fixed.  Any change in an ordinary branch under such a scan would signal inconsistent matching or numerics rather than a physical influence of the second exterior.

\begin{figure*}[!htp]
\centering
\includegraphics[width=\textwidth]{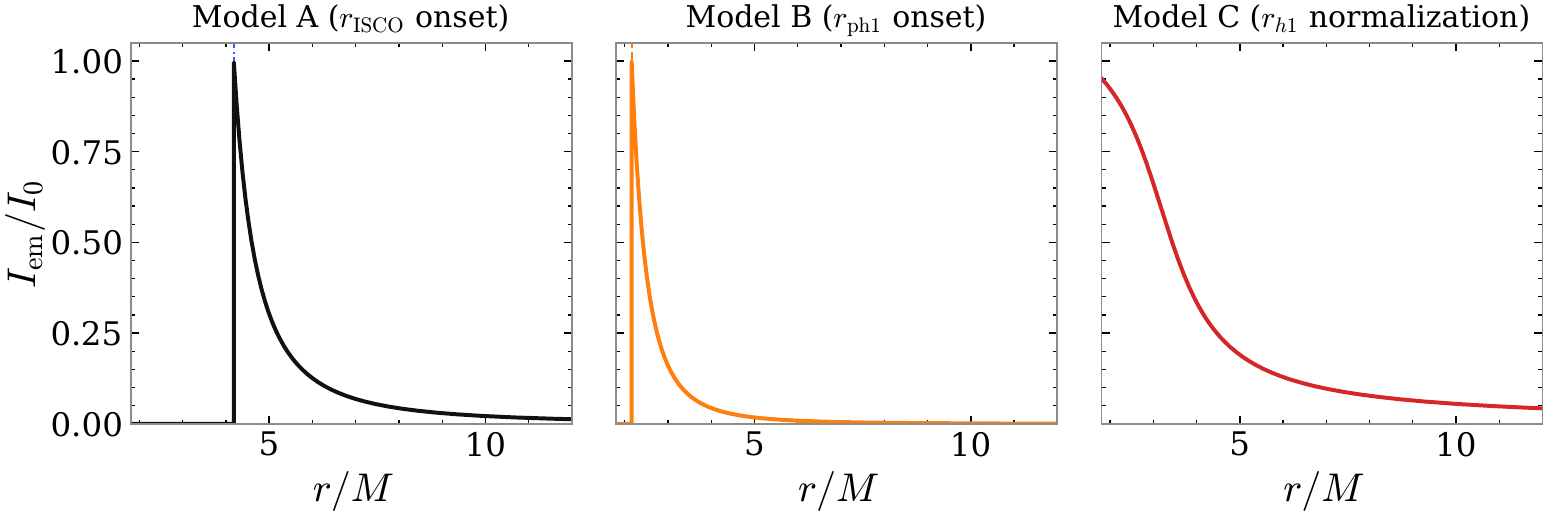}
\caption{Observer-side normalized bolometric emission profiles $I_{\rm em}/I_0$ as functions of the nonareal disk coordinate $r/M$ for Models A, B, and C.  The joined geometry is instantiated at $M_1=M_2\equiv M$, $\lambda/M^2=0.40$, $\bar c_q=0.08$, $w_q=-0.60$, and $a/M=1.80$, where $\bar c_q=c_q/M^{3w_q+1}$.  Although these values coincide with panel (g) of Fig.~\ref{fig:transfer-functions-matrix}, the plotted source laws depend only on the observer exterior $\Mone$.  The physical disk begins at the throat, which is the left frame boundary.  The relevant observer-side scales are $r_{h1}/M=1.35220779$, $r_{{\rm ph}1}/M=2.16071117$, and $r_{\rm ISCO}/M=4.18265897$.  The upper labels distinguish the sharp emission \emph{onsets} of Models A and B from the seed-horizon \emph{normalization} of Model C; $r_{h1}$ belongs to the excised seed region and is not an emitting edge of the joined disk.  All panels use $1.8\leq r/M\leq12$ and the same ordinate range.}
\label{fig:emission-profiles}
\end{figure*}

The innermost stable circular orbit is used as one phenomenological source scale.  For a unit-mass timelike particle in the observer exterior, define $D=F_1H'-F_1'H$.  Here $D$ has dimension $\mathsf{L}$, since $F_1$ is dimensionless and $H$ has dimension $\mathsf{L}^2$.  Circularity then gives
\begin{equation}
\begin{aligned}
 E_c^2=\frac{F_1^2H'}{D},
 \qquad &
 L_c^2=\frac{F_1'H^2}{D},
 \\
 \Omega_c^2&=\frac{F_1'}{H'}.
\end{aligned}
 \label{eq:circular-invariants}
\end{equation}
The quantities $E_c$, $L_c$, and $\Omega_c$ are, respectively, the specific energy, specific angular momentum, and coordinate angular velocity of the circular orbit.  Their dimensions are, respectively, $1$, $\mathsf{L}$, and $\mathsf{L}^{-1}$.  Physical circular motion on the selected branch requires positive $F_1$, $D$, $F_1'$, and $H'$.  Marginal radial stability is determined by
\begin{equation}
 F_1F_1''H'-2(F_1')^2H'
 +2F_1F_1'\frac{(H')^2}{H}
 -F_1F_1'H''=0.
 \label{eq:general-isco}
\end{equation}
This is the nonareal innermost-stable-circular-orbit (ISCO) condition.  It reduces to the familiar areal expression when $H=r^2$ and recovers $r_{\rm ISCO}=6M$ in the Schwarzschild limit.  For the configuration used in Fig.~\ref{fig:emission-profiles}, the physical root is $r_{\rm ISCO}/M=4.18265897$, outside both the throat and the observer-side photon sphere.  This value sets the Model-A source onset; it does not imply that the phenomenological emitters introduced below follow geodesic circular motion.

We adopt three bolometric source laws to test how radial support selects the same geometric transfer family.  Model A begins at the nonareal ISCO,
\begin{equation}
 \frac{\Iem^{(A)}(r)}{I_0}=
 \begin{cases}
 \left[\dfrac{M}{r-(r_{\rm ISCO}-M)}\right]^2,
 &r>r_{\rm ISCO},\\[4pt]
 0,&r\leq r_{\rm ISCO}.
 \end{cases}
 \label{eq:emission-a}
\end{equation}
Model B begins at the observer-side photon sphere,
\begin{equation}
 \frac{\Iem^{(B)}(r)}{I_0}=
 \begin{cases}
 \left[\dfrac{M}{r-(r_{{\rm ph}1}-M)}\right]^3,
 &r>r_{{\rm ph}1},\\[4pt]
 0,&r\leq r_{{\rm ph}1},
 \end{cases}
 \label{eq:emission-b}
\end{equation}
and Model C extends smoothly toward the inner seed region,
\begin{equation}
 \frac{\Iem^{(C)}(r)}{I_0}=
 \begin{cases}
 \dfrac{\frac{\pi}{2}-\arctan\!\left[\frac{r-(r_{\rm ISCO}-M)}{M}\right]}
 {\frac{\pi}{2}-\arctan\!\left[\frac{r_{h1}-(r_{\rm ISCO}-M)}{M}\right]},
 &r>r_{h1},\\[10pt]
 0,&r\leq r_{h1}.
 \end{cases}
 \label{eq:emission-c}
\end{equation}
The constant $I_0$ fixes a common intensity normalization.  Model A isolates intersections outside the timelike stability scale, Model B weights the neighborhood outside the null critical radius more strongly, and Model C tests a broader inner support.  In the joined spacetime, every formal source domain is intersected with $r\geq a$.  The horizon appearing in Model C is therefore only a normalization scale inherited from the unexcised pure-polymer seed, not an emitting surface available to the wormhole disk.  A transfer branch can consequently be geometrically present but dark in Models A or B if all of its retained intersections fall below the corresponding cutoff.

The terminology on the upper axes is physically deliberate.  In Models A and B, ``onset'' is a genuine boundary of the prescribed source support: Eqs.~\eqref{eq:emission-a} and \eqref{eq:emission-b} set the intensity to zero at and below $r_{\rm ISCO}$ and $r_{{\rm ph}1}$, respectively, while the exterior one-sided limits equal $I_0$.  The almost vertical rises are therefore prescribed emissivity discontinuities, not metric singularities or failures of the $4800$-point radial sampling.  Their meanings are nevertheless different.  The ISCO is fixed by the timelike marginal-stability condition in Eq.~\eqref{eq:general-isco}, whereas the photon sphere is the unstable null orbit defined by Eq.~\eqref{eq:photon-sphere-critical-impact}.  Since $r_{\rm ISCO}>r_{{\rm ph}1}>a$ here, Model A removes a wider inner portion of the physical disk than Model B.  Either cutoff is an illumination choice: disk intersections and null trajectories below it continue to exist geometrically but carry no weight in that model.

The third panel instead uses ``normalization'' because $r_{h1}$ enters Eq.~\eqref{eq:emission-c} only in the denominator.  In the unexcised seed geometry the chosen normalization would give $I_{\rm em}=I_0$ at that horizon, but the thin-shell construction removes this region and starts the physical disk at $a/M=1.80$.  The first accessible value is consequently evaluated at the throat and is approximately $0.952I_0$.  Model C therefore illuminates the whole retained disk, including the interval from the throat to the observer-side photon sphere, without assigning emission to the horizon or to the removed polymer interior.  The cosmological horizon on $\Mtwo$, located at $r_c/M\simeq20.82$ for this configuration, is likewise not a source boundary: it constrains propagation in the second exterior, whereas all three emission laws are defined on $\Mone$.  Moreover, $r$ is nonareal; by Eq.~\eqref{eq:angular-function}, the circumference radius is $\sqrt{H(r)}$, so horizontal coordinate separations in the figure are not proper radial distances.

The panels also encode three distinct large-radius weightings.  Model A decays quadratically beyond its ISCO edge, Model B decays cubically beyond its photon-sphere edge, and the arctangent prescription of Model C retains the slower inverse-radius tail.  The ordering is confirmed beyond the displayed interval by a finite numerical scan to $r/M=100$, where the normalized intensities are approximately $1.07\times10^{-4}$, $1.04\times10^{-6}$, and $3.91\times10^{-3}$ for Models A, B, and C, respectively.  These values document the computed finite range and are not an asymptotic proof.  Thus the same transfer radius receives substantially different weights even when all three models illuminate it: Model B suppresses the outer disk most rapidly, while Model C retains the broadest radial support.  Conversely, a transfer solution below $r_{\rm ISCO}$ is dark in Model A, one below $r_{{\rm ph}1}$ is dark in both Models A and B, and every retained solution with $r_m\geq a$ can contribute in Model C.

This support hierarchy can be compared point by point with panel (g) of Fig.~\ref{fig:transfer-functions-matrix}, because both calculations use $\lambda/M^2=0.40$, $\bar c_q=0.08$, and $w_q=-0.60$.  The direct and ordinary higher-order transfer functions are common geometric solutions for all three source models, whereas the displayed returned third-order branch owes its existence to the excursion through $\Mtwo$.  Applying Eqs.~\eqref{eq:emission-a}--\eqref{eq:emission-c} to those radii can suppress, reveal, or reweight a branch, but cannot displace it or generate a new null trajectory.  In particular, the portion of a returned branch emerging close to the throat is removed by both sharp-cutoff models; Model B begins to illuminate it once its transfer radius exceeds $r_{{\rm ph}1}$, Model A only after the larger ISCO threshold is crossed, and Model C weights its full retained support.  Branch segments above the plotted $r_m/M=6$ ceiling of Fig.~\ref{fig:transfer-functions-matrix} remain part of this comparison because that ceiling is visual rather than an integration cutoff.  Whether any weighted segment produces a resolved image feature still requires the observed-intensity calculation.

The absence of $\bar c_q$ and $w_q$ families in Fig.~\ref{fig:emission-profiles} is an exact control rather than an omitted scan.  These parameters modify only $F_2$, while the characteristic radii and source laws plotted here are constructed entirely from $F_1$ and $H$ on the observer side.  Varying quintessence at fixed $(M,\lambda,a)$ may reorganize the returned transfer branches or send a transmitted ray toward the cosmological horizon, but it leaves all three curves in this figure unchanged.  In contrast, changing $\lambda$ deforms both $F_1$ and $H$, moves $r_{{\rm ph}1}$ and $r_{\rm ISCO}$, changes the seed-horizon normalization in Model C, and therefore defines a genuinely different source profile.  Holding the polymer scale fixed here is essential for attributing later seed--thin-shell intensity differences to the propagation geometry rather than to a simultaneous change in illumination.

Finally, Eq.~\eqref{eq:observed-intensity} converts emitted into received brightness by evaluating the selected profile at the ordered transfer radii, weighting every contribution by $F_1(r_m)^2$, and then summing the image orders.  Consequently, the substantial Model-C emissivity near the throat may be strongly redshifted, whereas the discontinuous Model-A or Model-B onset can generate a sharp screen feature when a transfer branch crosses the corresponding source edge.  Critical compression, branch demagnification, overlapping orders, and instrumental response remain distinct layers of the image formation.  Figure~\ref{fig:emission-profiles} therefore establishes three controlled illumination prescriptions that expose how the same transfer geometry is reweighted in the like-for-like seed and thin-shell comparison.

For the initial radiative model the emitters are static in $\Mone$, the observer is static at its normalized infinity, the disk is face-on, geometrically and optically thin, and emission is isotropic and bolometric.  The frequency shift for the $m$th disk intersection is $g_m=\sqrt{F_1(r_m)}$.  Liouville invariance then gives
\begin{equation}
\begin{aligned}
 \Iobs(b_1)&=\sum_{m=1}^{3}g_m^4\,
 \Iem\!\left(r_m(b_1)\right)
 \\
 &=\sum_{m=1}^{3}F_1(r_m)^2\,
 \Iem\!\left(r_m(b_1)\right).
\end{aligned}
 \label{eq:observed-intensity}
\end{equation}
This is a discrete sum over the first three transfer orders, not a volumetric line-of-sight integral.  The factor $F_1(r_m)^2$ weights a pre-existing branch; it cannot create a new geodesic solution or move a critical edge.  If the emitters are later placed on circular orbits, the redshift must be rederived to include their motion and the associated directional Doppler factor.

Axial symmetry gives the face-on map $\mathscr{I}(x_s,y_s)=\Iobs(\sqrt{x_s^2+y_s^2})$, with $(x_s,y_s)$ measured on the observer screen.  Thus every finite onset, narrow maximum, and broad tail of the radial profile becomes a circular feature in the image.  A central brightness depression is not automatically a geometric shadow: it can combine capture or cross-throat loss, gravitational redshift, the shell truncation of the disk, and the support of the chosen emission law.

The radial intensity curve, full image, and local image must be evaluated from the same accepted transfer data and radial profile.  Near $b_{c1}$, $Zb_{c2}$, and any emission-support crossing, dedicated refinement is required because a logarithmic critical enhancement and a finite source-edge onset have different physical origins.  A sharp feature is retained only after its one-sided value and transfer order have been identified and its peak height and location are stable under refinement.  These numerical statements are part of the interpretation of Eq.~\eqref{eq:observed-intensity}: without them, a plotted narrow ring cannot be distinguished reliably from under-resolution.

For the converged profiles and maps presented below, the selected parameter slice establishes a finite set of ordinary and returned intensity features in the static, face-on model.  The common source laws, first three transfer orders, and unconvolved bolometric intensities provide a controlled basis for attributing each change in morphology to the matched propagation channel and its emitting support.  This intrinsic representation preserves the location, order, and source dependence of the narrow structures before additional radiative or instrumental processing is introduced.

\begin{figure}[!htp]
\centering
\resizebox{1\linewidth}{!}{%
{\setlength{\tabcolsep}{1.2pt}
\renewcommand{\arraystretch}{0}
\begin{tabular}{@{}ccc@{}}
\includegraphics[width=0.4267\linewidth,trim=0 26.5bp 0 2.88bp,clip]{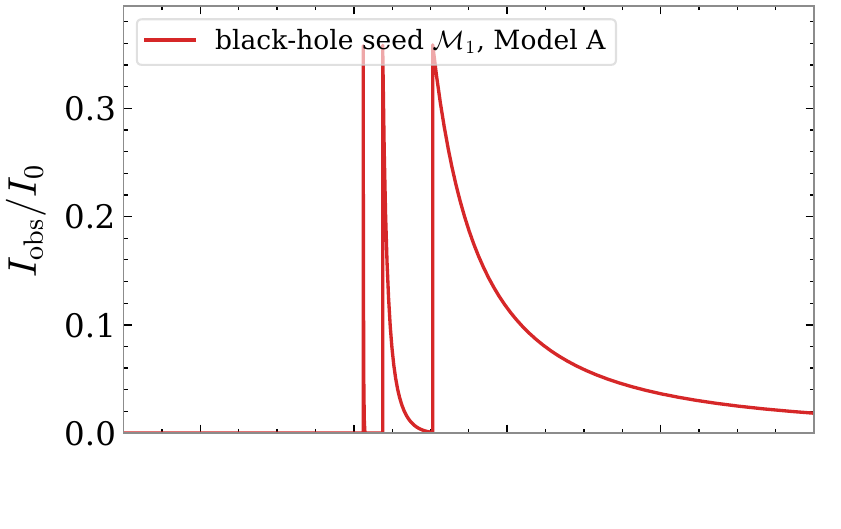} &
\raisebox{0.0204\linewidth}{\includegraphics[width=0.216\linewidth]{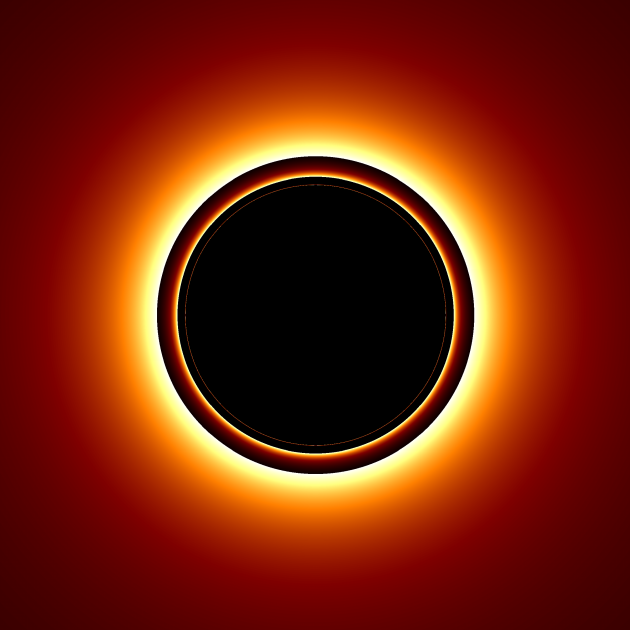}} &
\raisebox{0.0204\linewidth}{\includegraphics[width=0.216\linewidth]{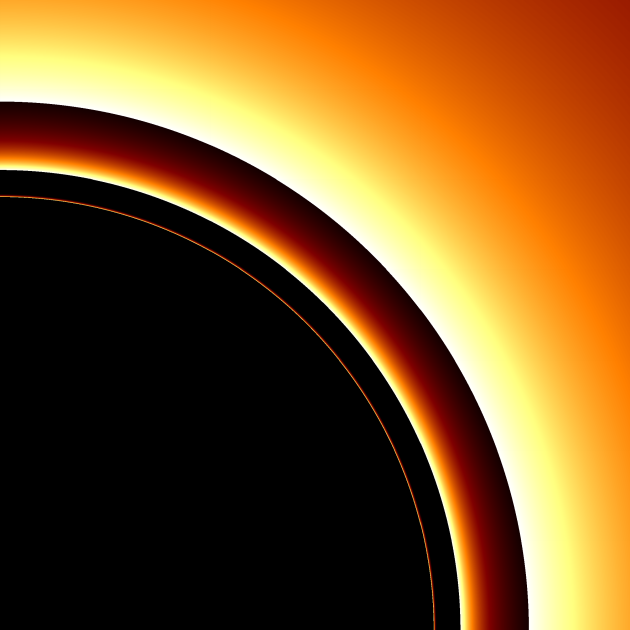}} \\[-7pt]
\includegraphics[width=0.4267\linewidth,trim=0 26.5bp 0 2.88bp,clip]{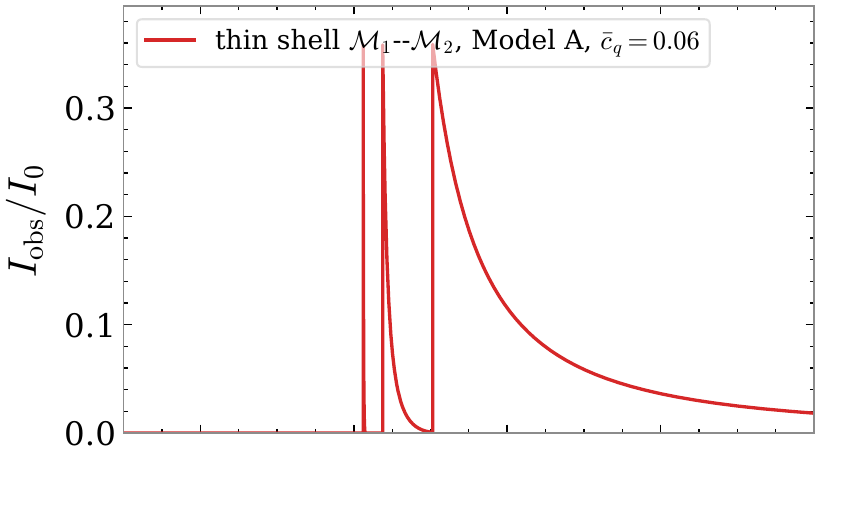} &
\raisebox{0.0204\linewidth}{\includegraphics[width=0.216\linewidth]{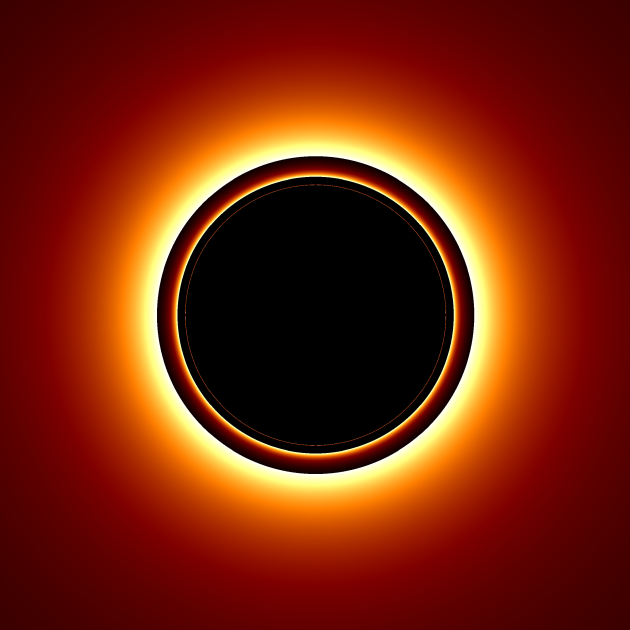}} &
\raisebox{0.0204\linewidth}{\includegraphics[width=0.216\linewidth]{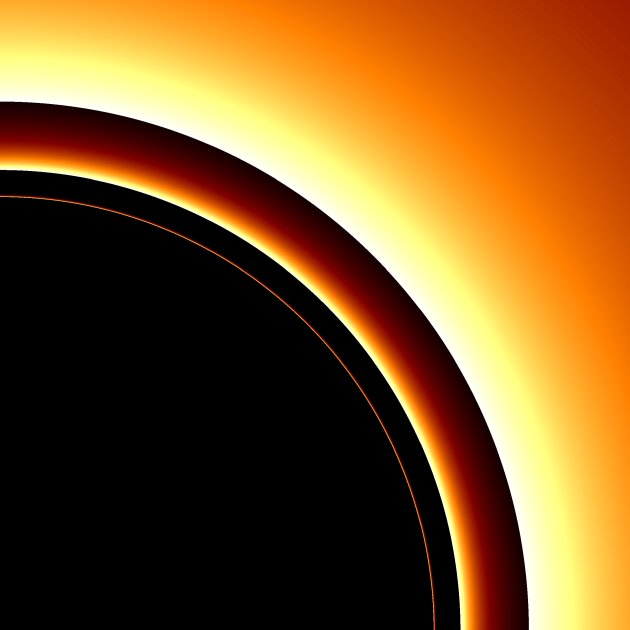}} \\[-7pt]
\includegraphics[width=0.4267\linewidth,trim=0 26.5bp 0 2.88bp,clip]{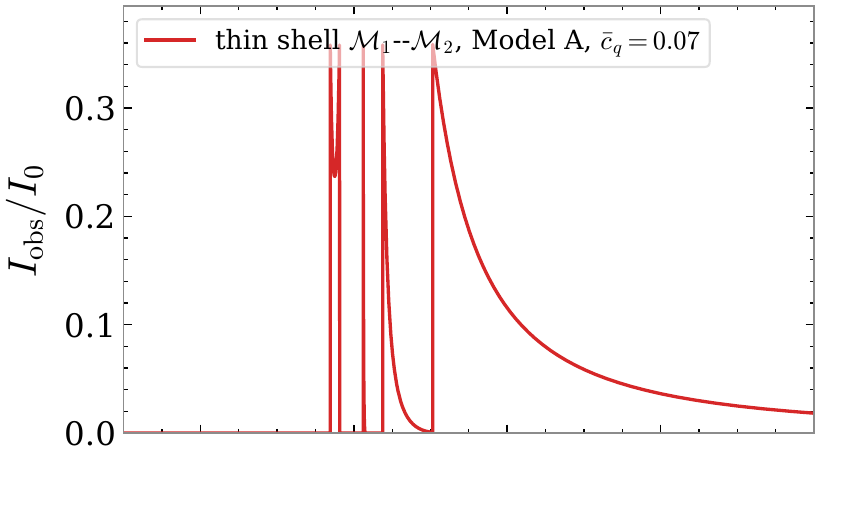} &
\raisebox{0.0204\linewidth}{\includegraphics[width=0.216\linewidth]{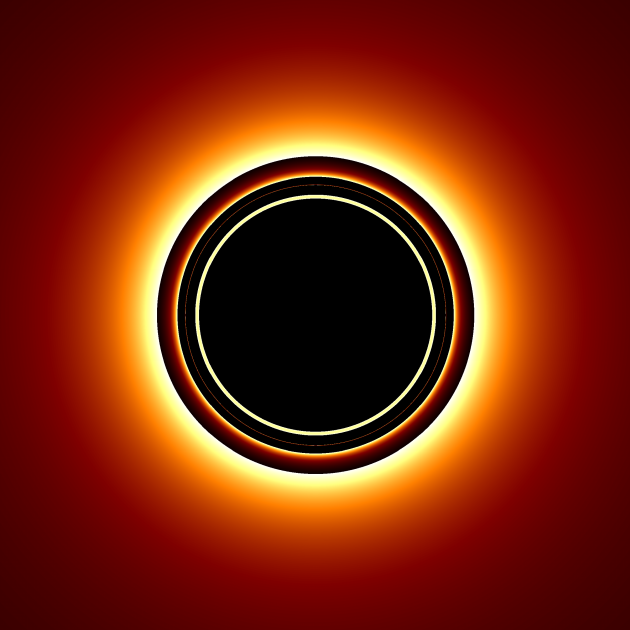}} &
\raisebox{0.0204\linewidth}{\includegraphics[width=0.216\linewidth]{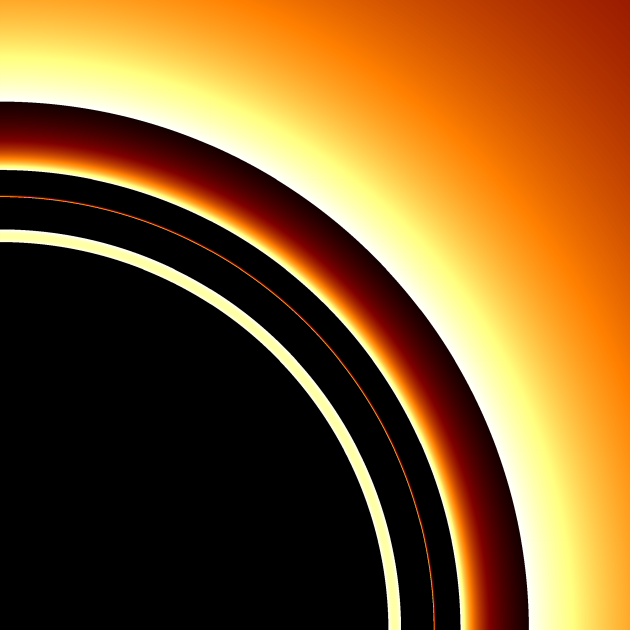}} \\[-7pt]
\includegraphics[width=0.4267\linewidth,trim=0 26.5bp 0 2.88bp,clip]{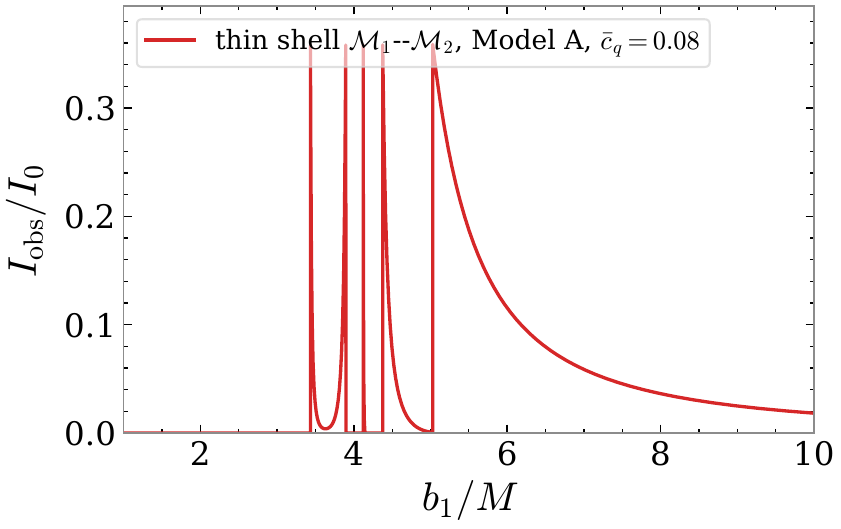} &
\raisebox{0.0204\linewidth}{\includegraphics[width=0.216\linewidth]{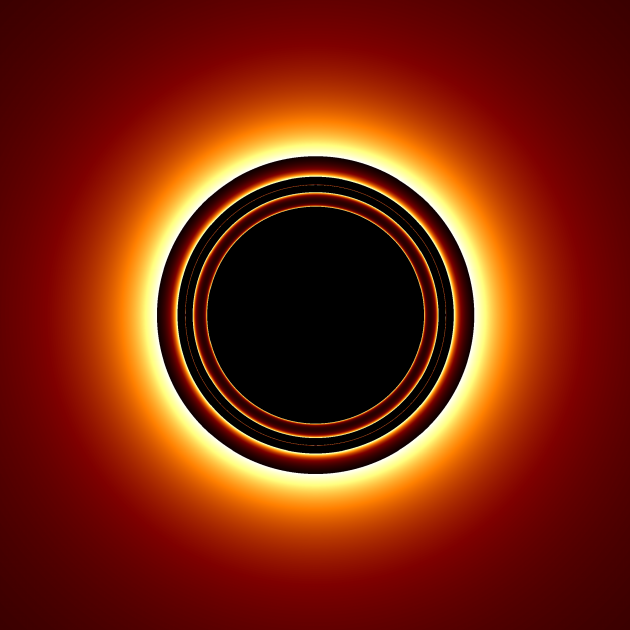}} &
\raisebox{0.0204\linewidth}{\includegraphics[width=0.216\linewidth]{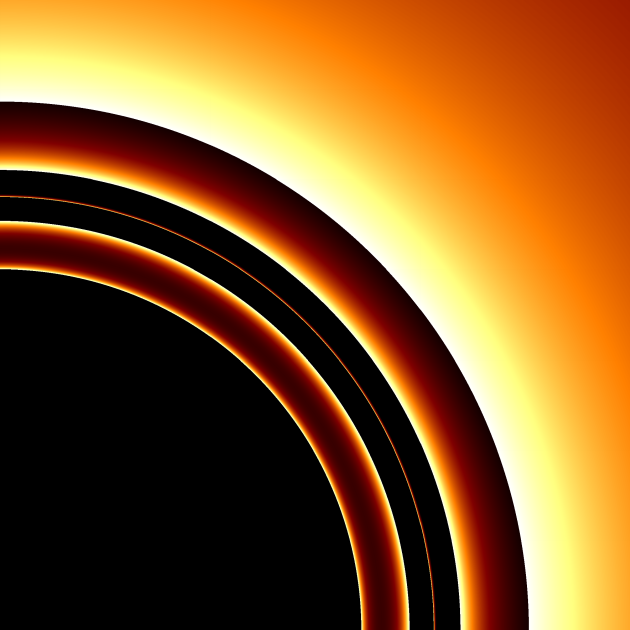}}
\end{tabular}}%
}
\caption{Observed-intensity profiles and face-on optical maps for the ISCO-truncated emission Model A.  The common geometry has $M=M_1=M_2=1$ and $\lambda/M^2=0.40$.  The upper row is the pure-polymer black-hole seed; the lower rows are matched thin shells with $a/M=1.80$, $w_q=-0.60$, and, from top to bottom, $\bar c_q=0.06$, $0.07$, and $0.08$, where $\bar c_q=c_q/M^{3w_q+1}$.  Columns show $I_{\rm obs}/I_0$, its full face-on map, and a first-quadrant enlargement.  Both maps in each row are generated from the plotted profile, and all rows share one intensity normalization.  The profiles cover $1\leq b_1/M\leq10$; the full and enlarged maps span $[-10M,10M]^2$ and $[0,6M]^2$, respectively.  Color represents optical intensity, not matter density.}
\label{fig:iobs-density-local-model-a}
\end{figure}

The seed--thin-shell comparison for the ISCO-truncated emission prescription is shown in Fig.~\ref{fig:iobs-density-local-model-a}.

The Model-A matrix separates the invariant observer-side image family from the part of the signal generated by cross-throat propagation.  Model A illuminates only $r>r_{\rm ISCO}$, with $r_{\rm ISCO}/M=4.18265897$, and the black-hole control has a converged dominant maximum $I_{\rm obs}/I_0\simeq0.35877$ at $b_1/M\simeq5.02890$.  Because the source law, $F_1$, $H$, and the observer-side redshift are common to all four rows, this ordinary maximum is unchanged by $\bar c_q$.  Its invariance controls the one-sided part of the calculation; it does not imply equality of the complete profiles or of their integrated fluxes.

The quintessence dependence enters only after a returned intersection reaches the emitting support.  An independent minimization of the accumulated-angle condition, followed by bracketing checks on both sides of the root, places the first nonzero departure from the seed at $\bar c_q\simeq0.0689482$: at this value the returned third-order branch first reaches $r_{\rm ISCO}$, at $b_1/M\simeq3.76172$.  The $\bar c_q=0.06$ row is therefore still physically coincident with the seed under Model A, whereas $0.07$ is the first displayed row above the onset and $0.08$ lies farther inside the emissive regime.  Correspondingly, the first returned structures in the latter two rows occur near $b_1/M\simeq3.69$ and $3.44$.  Their inward migration follows the decrease of $Zb_{c2}$ and the widening of the returned interval; it is a change in propagation and source support, not a displacement of the observer-side photon sphere.

The full maps and their local enlargements are two renderings of the same face-on profile defined by Eq.~\eqref{eq:observed-intensity}.  The seed and $\bar c_q=0.06$ rows are dominated by the same ordinary annuli, while the $0.07$ and $0.08$ rows acquire additional concentric features as progressively more of the returned branch lies above the ISCO cutoff.  A returned null path can therefore be geometrically present and yet remain dark in this emission model.  Conversely, the surviving inner rings are jointly selected by the matched trajectory and the source edge; they do not signal an additional photon sphere or an independent emitting component.

The central depression likewise has mixed origins.  In the seed geometry it combines horizon capture, redshift, and the source cutoff; in the shell geometries, cross-throat transmission and reflection replace part of that causal bookkeeping.  Since the interval $a\leq r\leq r_{\rm ISCO}$ is dark in Model A, the central gap cannot be assigned to geometry alone.  Within $1\leq b_1/M\leq10$, the matrix shows that increasing $\bar c_q$ beyond the emission-support threshold exposes a larger fraction of the returned optical topology while preserving the ordinary peak.  The additional narrow structures consequently isolate the combined action of cross-throat propagation and the ISCO source edge.

\begin{figure}[!htp]
\centering
\resizebox{1\linewidth}{!}{%
{\setlength{\tabcolsep}{1.2pt}
\renewcommand{\arraystretch}{0}
\begin{tabular}{@{}ccc@{}}
\includegraphics[width=0.4267\linewidth,trim=0 26.5bp 0 2.88bp,clip]{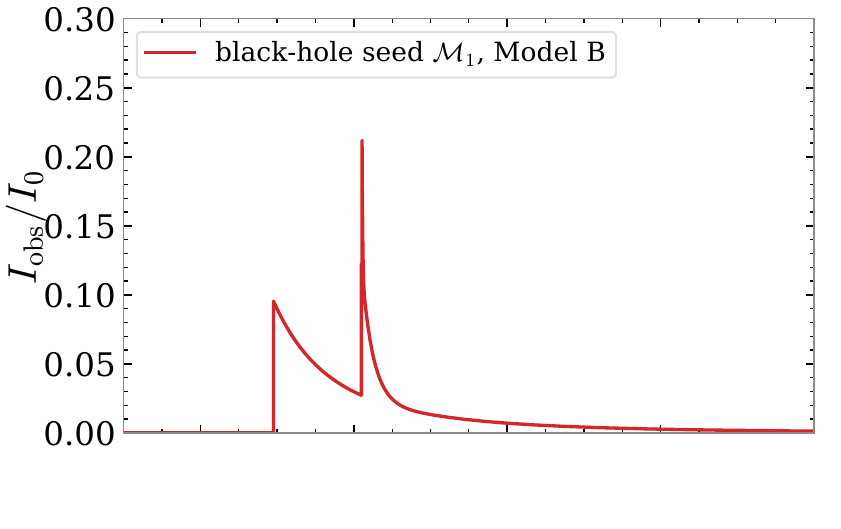} &
\raisebox{0.0204\linewidth}{\includegraphics[width=0.216\linewidth]{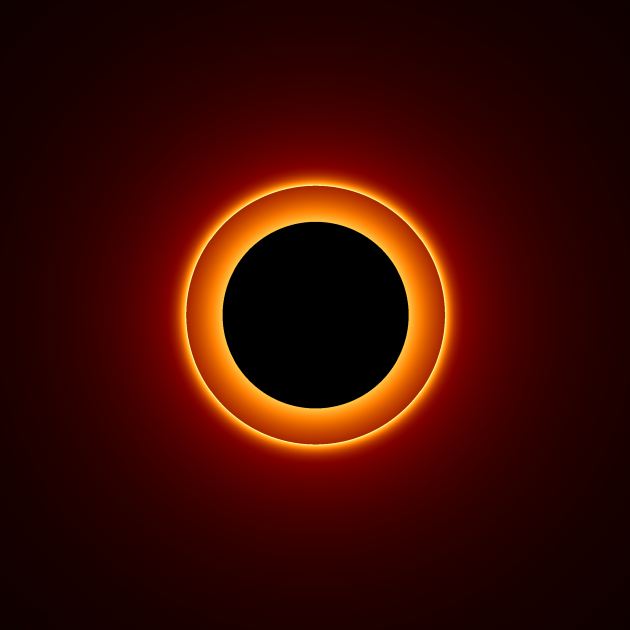}} &
\raisebox{0.0204\linewidth}{\includegraphics[width=0.216\linewidth]{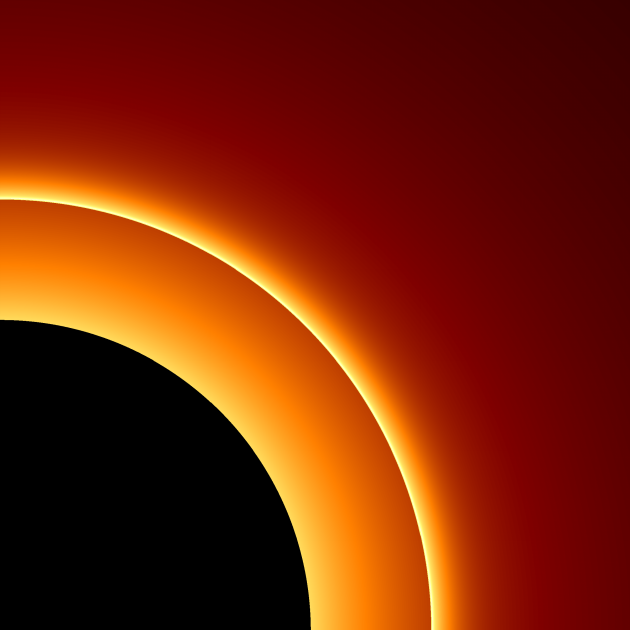}} \\[-7pt]
\includegraphics[width=0.4267\linewidth,trim=0 26.5bp 0 2.88bp,clip]{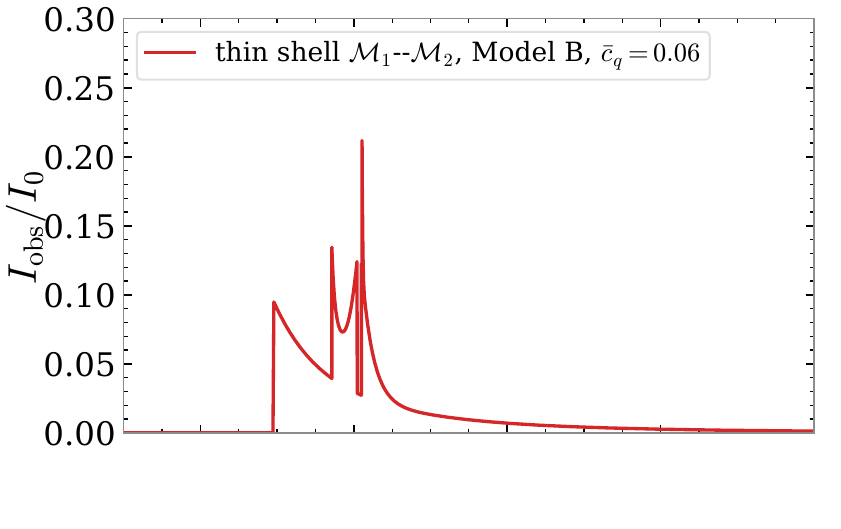} &
\raisebox{0.0204\linewidth}{\includegraphics[width=0.216\linewidth]{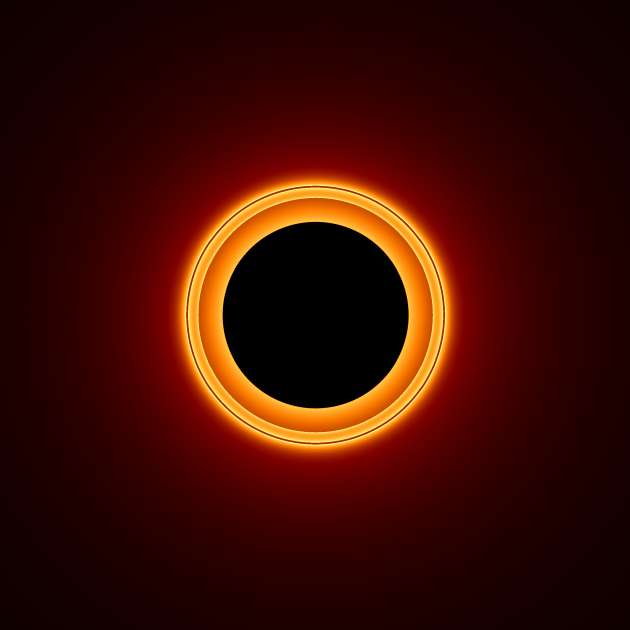}} &
\raisebox{0.0204\linewidth}{\includegraphics[width=0.216\linewidth]{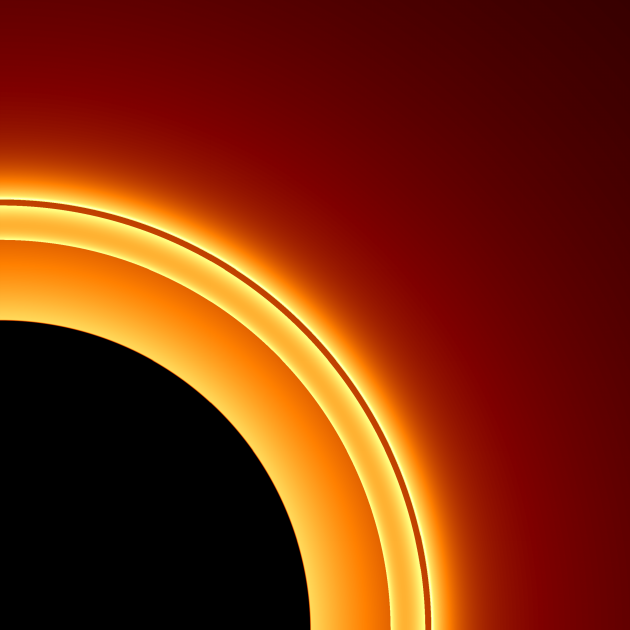}} \\[-7pt]
\includegraphics[width=0.4267\linewidth,trim=0 26.5bp 0 2.88bp,clip]{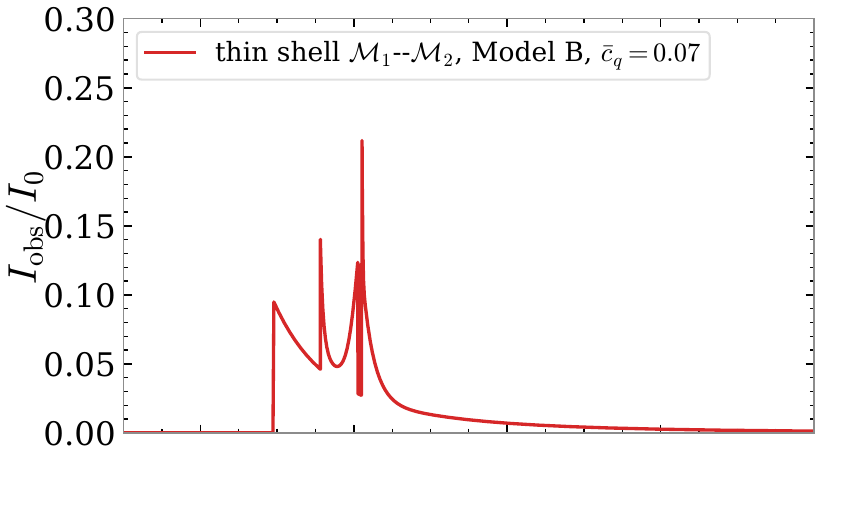} &
\raisebox{0.0204\linewidth}{\includegraphics[width=0.216\linewidth]{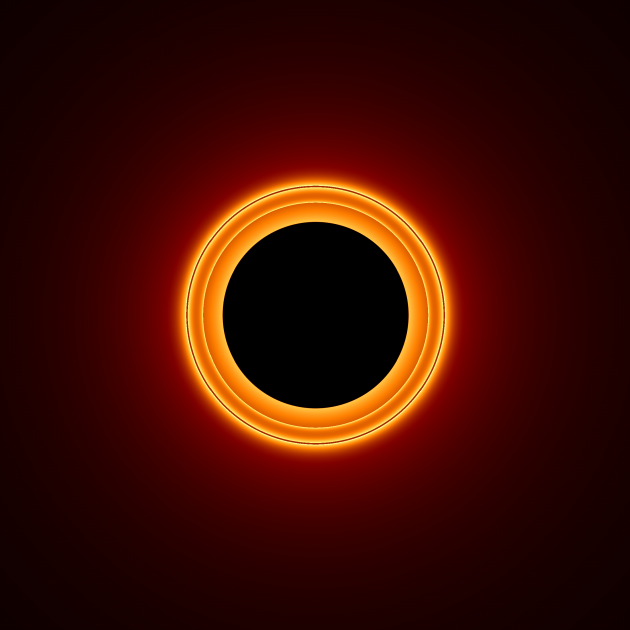}} &
\raisebox{0.0204\linewidth}{\includegraphics[width=0.216\linewidth]{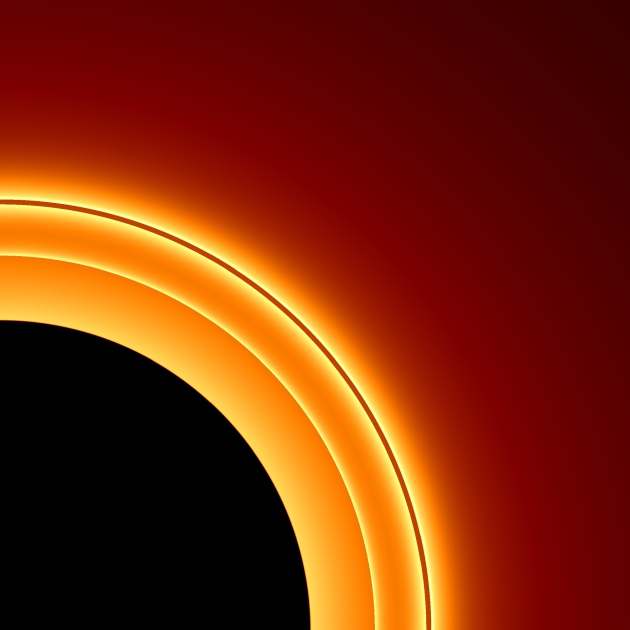}} \\[-7pt]
\includegraphics[width=0.4267\linewidth,trim=0 26.5bp 0 2.88bp,clip]{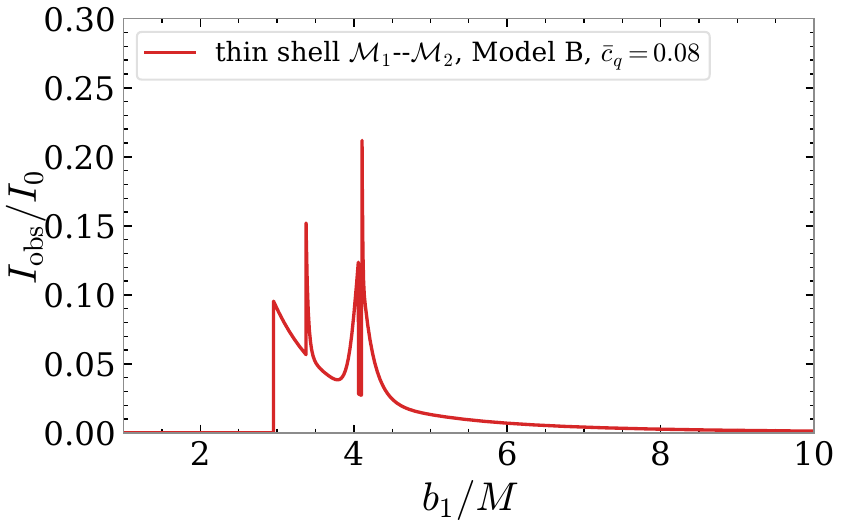} &
\raisebox{0.0204\linewidth}{\includegraphics[width=0.216\linewidth]{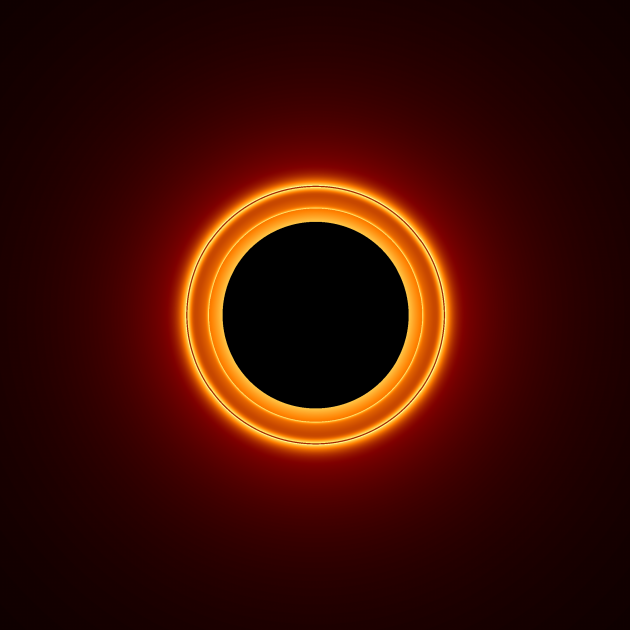}} &
\raisebox{0.0204\linewidth}{\includegraphics[width=0.216\linewidth]{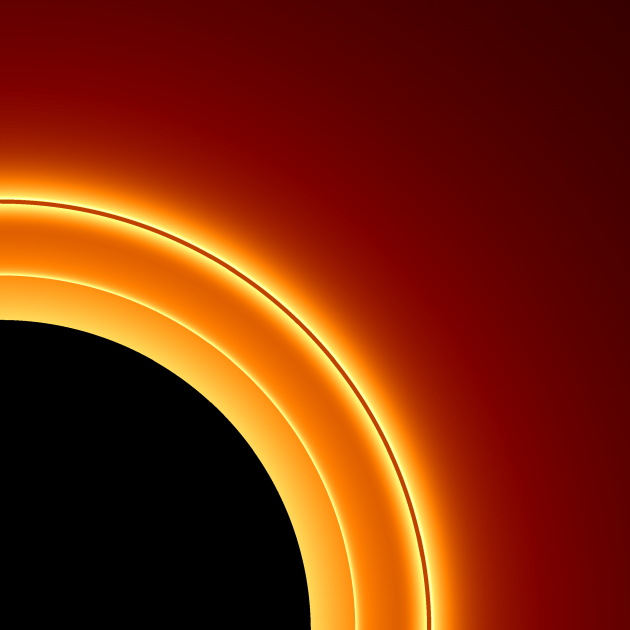}}
\end{tabular}}%
}
\caption{Observed-intensity profiles and face-on optical maps for the photon-sphere-truncated emission Model B.  The common geometry has $M=M_1=M_2=1$ and $\lambda/M^2=0.40$.  The upper row is the pure-polymer black-hole seed; the lower rows are matched thin shells with $a/M=1.80$, $w_q=-0.60$, and, from top to bottom, $\bar c_q=0.06$, $0.07$, and $0.08$.  Columns show $I_{\rm obs}/I_0$, its full face-on map, and a first-quadrant enlargement.  Both maps in each row are generated from the plotted profile, and all rows share one intensity normalization.  The profiles cover $1\leq b_1/M\leq10$ and $0\leq I_{\rm obs}/I_0\leq0.3$; the full and enlarged maps span $[-10M,10M]^2$ and $[0,6M]^2$, respectively.  Color represents optical intensity, not matter density.}
\label{fig:iobs-density-local-model-b}
\end{figure}

Replacing the ISCO cutoff by the photon-sphere cutoff produces the Model-B comparison displayed in Fig.~\ref{fig:iobs-density-local-model-b}.

Model B begins at the observer-side photon sphere, $r_{{\rm ph}1}/M=2.16071117$, and therefore illuminates a larger retained interval than Model A.  Its ordinary critical component is fixed by $\Mone$ and converges to $I_{\rm obs}/I_0\simeq0.21184$ near $b_1/M=4.10795$, essentially the observer-side critical impact $b_{c1}/M=4.10798$.  This component is common to the seed and shell rows.  The additional finite-impact peaks arise only after a ray crosses the throat, turns in $\Mtwo$, returns to $\Mone$, and intersects the supported part of the disk; their positions and multiplicity, rather than the invariant ordinary maximum, encode the asymmetry.

The onset occurs at a smaller quintessence amplitude than in Model A because the support edge is the photon sphere rather than the larger ISCO radius.  The same two-sided root test gives $\bar c_q\simeq0.0427948$, where the returned third-order branch first reaches $r_{{\rm ph}1}$ at $b_1/M\simeq3.96669$.  All three displayed shell rows are therefore above the Model-B threshold, explaining why $\bar c_q=0.06$ already differs from the seed.  Increasing $\bar c_q$ moves $Zb_{c2}$ inward and broadens the returned screen interval, but it does not force every local peak to grow monotonically: the matching factor, the second-side turning integral, and the location at which a branch crosses the discontinuous emissivity edge vary together.

The full images and local enlargements translate the same radial profiles into circular morphology.  The seed row is dominated by the ordinary critical annulus, whereas the shell rows add narrow inner annuli at the accepted returned intersections.  These enlargements separate the inner features from the broad outer brightness without introducing a second source.  A geometrically allowed branch below $r_{{\rm ph}1}$ remains dark under Eq.~\eqref{eq:emission-b}; a visible inner ring therefore requires both an admissible returned path and an intersection above the source edge.

The sharp peaks combine two mechanisms that must remain distinct.  Critical compression near $b_{c1}$ enhances the ordinary ring continuously, whereas the piecewise source law creates a finite onset when a transfer radius crosses $r_{{\rm ph}1}$.  The adaptive profiles resolve the narrow returned structures and their peak surface brightness before instrumental convolution.  Within the sampled gate and source model, the matrix therefore establishes a quintessence-driven change in branch multiplicity and inner morphology whose onset is fixed by the photon-sphere source edge.

\begin{figure}[!htp]
\centering
\resizebox{1\linewidth}{!}{%
{\setlength{\tabcolsep}{1.2pt}
\renewcommand{\arraystretch}{0}
\begin{tabular}{@{}ccc@{}}
\includegraphics[width=0.4267\linewidth,trim=0 26.5bp 0 2.88bp,clip]{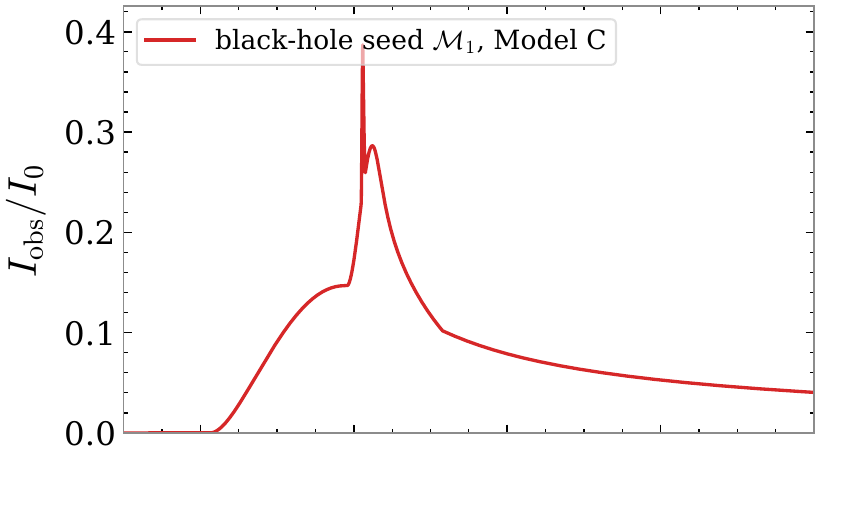} &
\raisebox{0.0204\linewidth}{\includegraphics[width=0.216\linewidth]{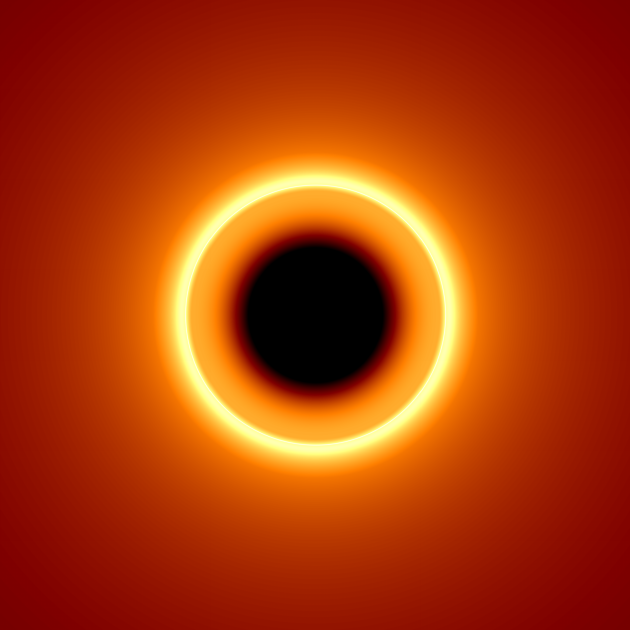}} &
\raisebox{0.0204\linewidth}{\includegraphics[width=0.216\linewidth]{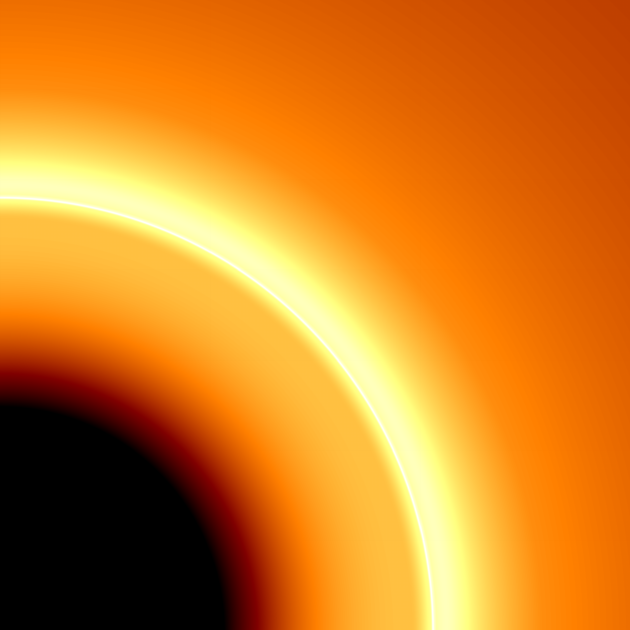}} \\[-7pt]
\includegraphics[width=0.4267\linewidth,trim=0 26.5bp 0 2.88bp,clip]{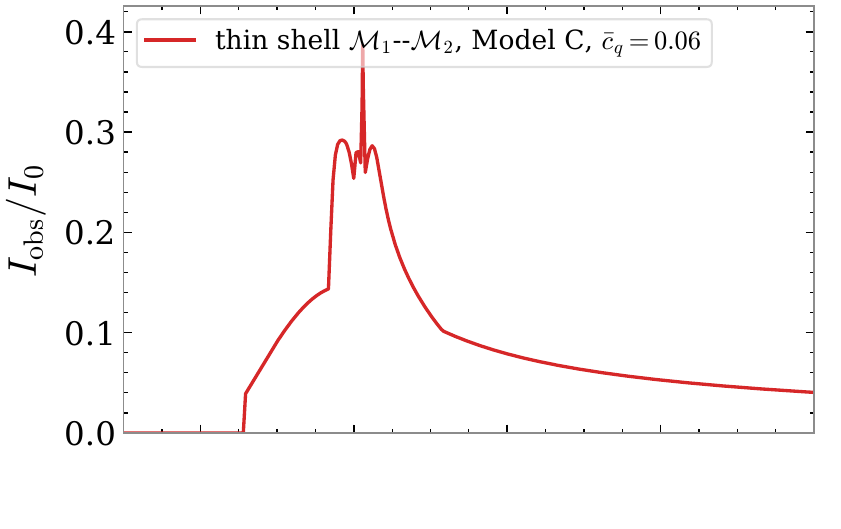} &
\raisebox{0.0204\linewidth}{\includegraphics[width=0.216\linewidth]{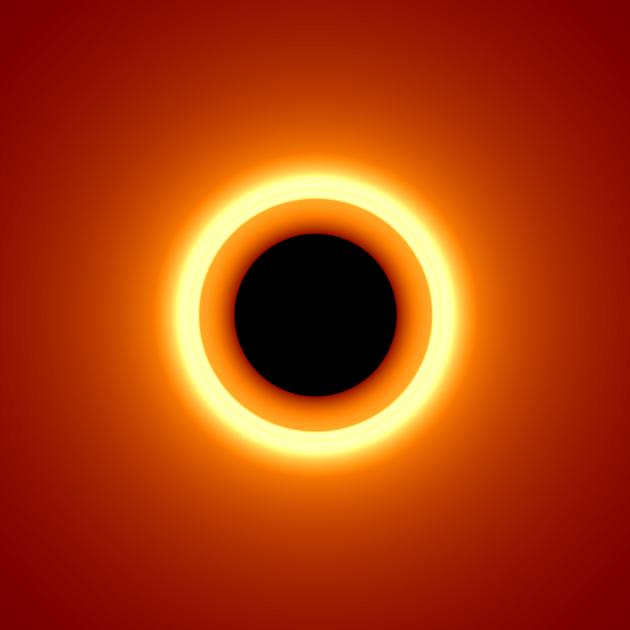}} &
\raisebox{0.0204\linewidth}{\includegraphics[width=0.216\linewidth]{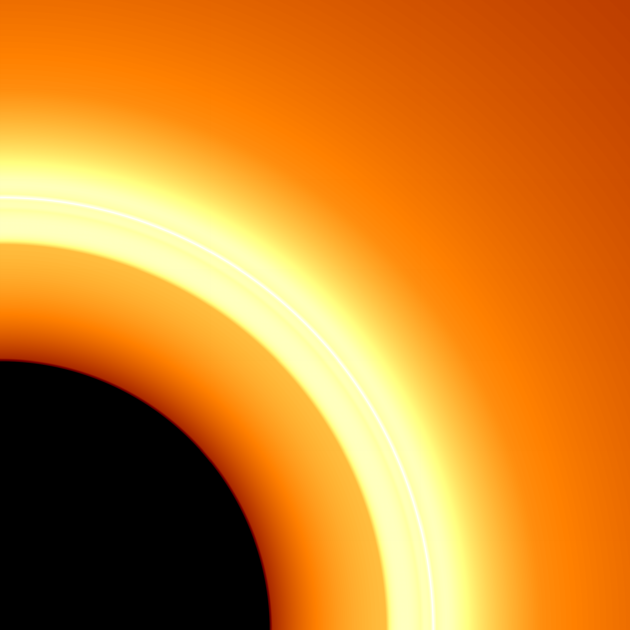}} \\[-7pt]
\includegraphics[width=0.4267\linewidth,trim=0 26.5bp 0 2.88bp,clip]{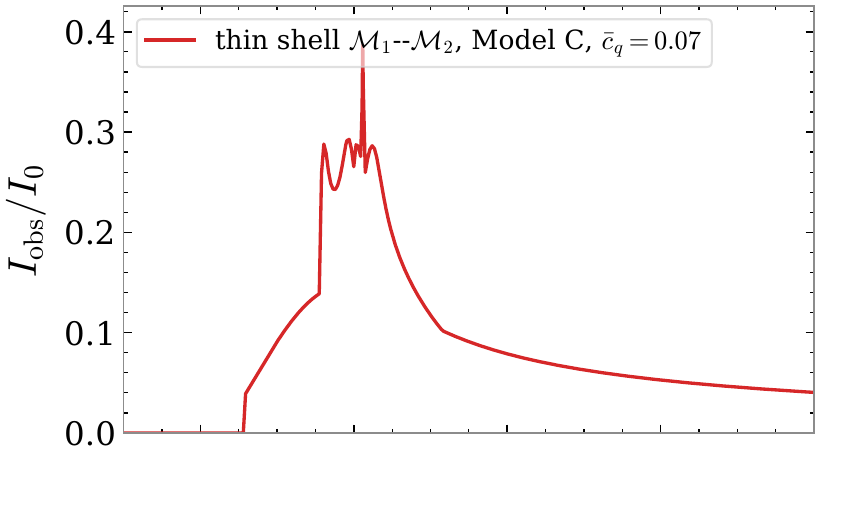} &
\raisebox{0.0204\linewidth}{\includegraphics[width=0.216\linewidth]{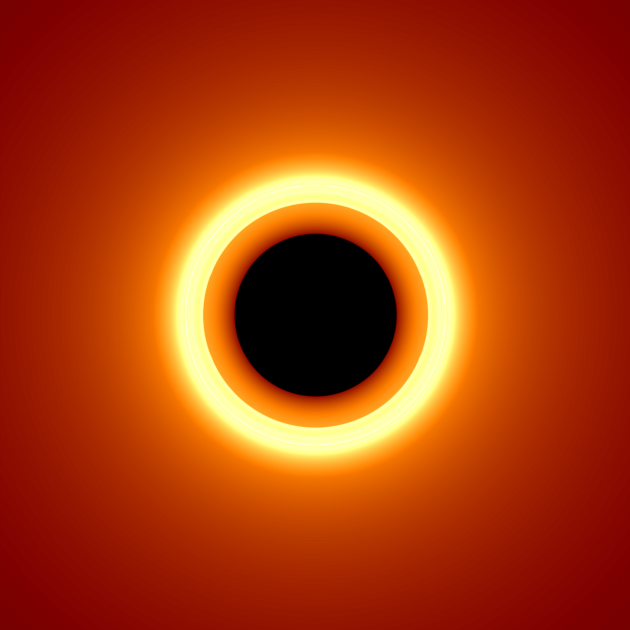}} &
\raisebox{0.0204\linewidth}{\includegraphics[width=0.216\linewidth]{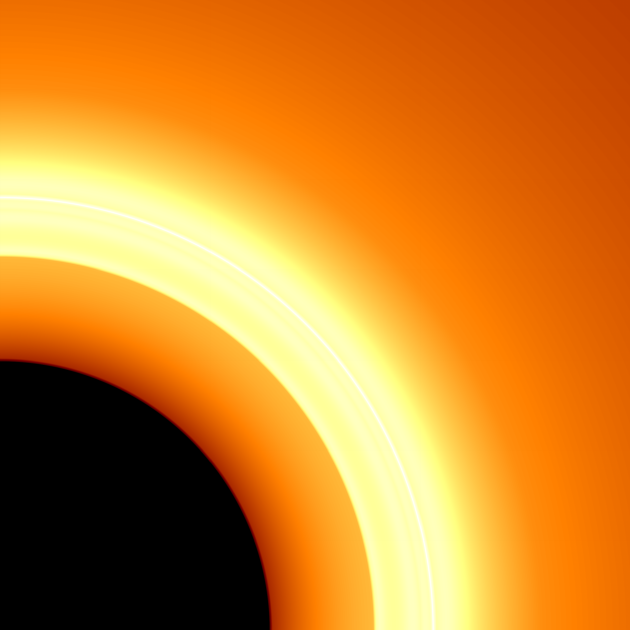}} \\[-7pt]
\includegraphics[width=0.4267\linewidth,trim=0 26.5bp 0 2.88bp,clip]{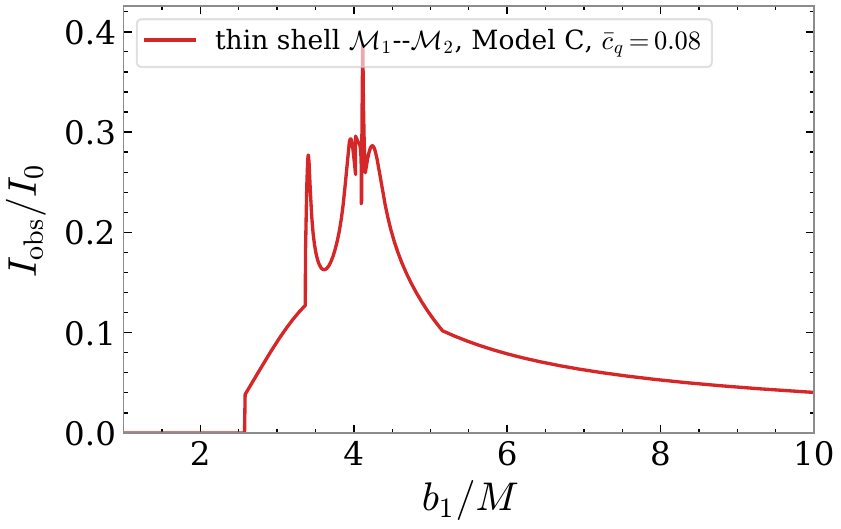} &
\raisebox{0.0204\linewidth}{\includegraphics[width=0.216\linewidth]{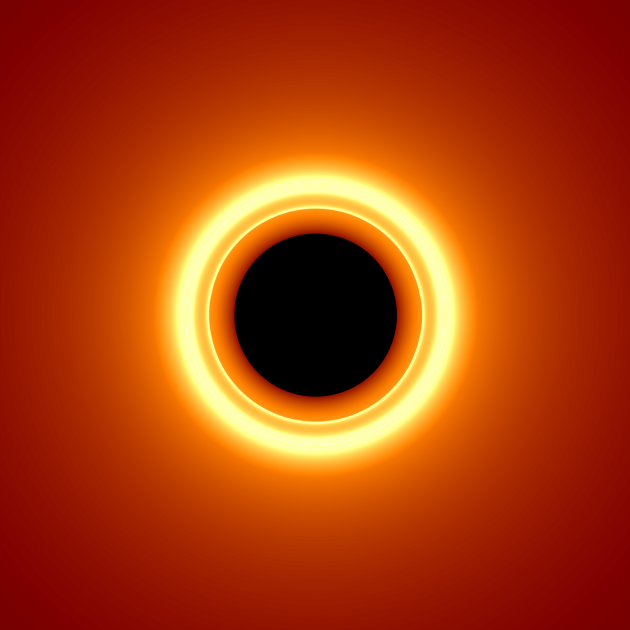}} &
\raisebox{0.0204\linewidth}{\includegraphics[width=0.216\linewidth]{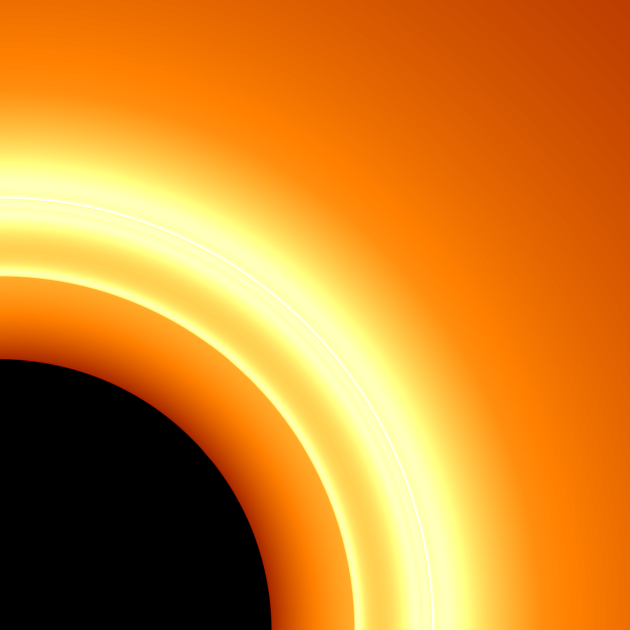}}
\end{tabular}}%
}
\caption{Observed-intensity profiles and face-on optical maps for the broad-support emission Model C.  The common geometry has $M=M_1=M_2=1$ and $\lambda/M^2=0.40$.  The upper row is the pure-polymer black-hole seed; the lower rows are matched thin shells with $a/M=1.80$, $w_q=-0.60$, and, from top to bottom, $\bar c_q=0.06$, $0.07$, and $0.08$.  For these shell rows, the emitting domain extends down to the throat.  Columns show $I_{\rm obs}/I_0$, its full face-on map, and a first-quadrant enlargement.  Both maps in each row are generated from the plotted profile, and all rows share one intensity normalization.  The profiles cover $1\leq b_1/M\leq10$; the full and enlarged maps span $[-10M,10M]^2$ and $[0,6M]^2$, respectively.  Color represents optical intensity, not matter density.}
\label{fig:iobs-density-local-model-c}
\end{figure}

The corresponding broad-support comparison, for which the thin-shell disk remains emissive down to the throat, is presented in Fig.~\ref{fig:iobs-density-local-model-c}.

Model C differs qualitatively from the two thresholded prescriptions because Eq.~\eqref{eq:emission-c} is nonzero throughout the retained disk and uses $r_{h1}$ only as a normalization scale.  The seed horizon is not an emitting surface of the joined spacetime; the thin-shell disk begins at $r=a$.  The ordinary observer-side profile is therefore common to all four rows, with a converged maximum $I_{\rm obs}/I_0\simeq0.38699$ at $b_1/M\simeq4.11641$.  Returned features remain emissive even when their disk intersections lie between the throat and $r_{{\rm ph}1}$, where Models A and B impose a dark interval.

The low-impact onset in the shell profiles has a direct geometric explanation.  The independently refined root $\Phi_1(b_1)=\pi/2$, with $\Phi_1(b_1)=\mathcal A_1(a,\infty;b_1)$ as defined in Eq.~\eqref{eq:second-side-angular-branches}, occurs at $b_1/M=2.5832965$.  Below it, the ray reaches the throat before the first observer-side disk target and then follows the transmitted channel toward the second-side cosmological horizon, so the one-disk model gives $I_{\rm obs}=0$.  Immediately above the root, the first intersection is born at $r=a$, where Model C has $I_{\rm em}/I_0\simeq0.95225$; after redshift weighting, the one-sided intensity tends to the finite value $I_{\rm obs}/I_0\simeq0.03833$.  This is a physical branch onset rather than a numerical truncation.  The root is fixed by $F_1$, $H$, and the throat, and is therefore independent of $\bar c_q$ at the present gate.  The first positive rendered samples for the $0.06$ and $0.07$ rows lie within about $7\times10^{-3}M$ of the refined root, so the quoted value should be taken from the root calculation rather than inferred from the pixel position of the jump.

The first thin-shell departure from the seed occurs at a much smaller amplitude, $\bar c_q\simeq0.00111610$.  Here the mechanism is not the crossing of a dark source edge: the mapped second-side critical boundary first opens an emission-supported third-order interval relative to the seed, whose corresponding screen limits are $b_1/M\simeq4.09617$ and $4.10341$.  All three displayed shell rows are therefore well above this onset.  Increasing $\bar c_q$ then changes $Z$, the second-side photon barrier, and the cosmological horizon, shifting and multiplying the returned annuli while leaving the broad ordinary maximum fixed.  Because Model C illuminates the interval $a<r<r_{{\rm ph}1}$, these inner channels remain visible when the same transfer radii would be suppressed by Models A and B.

The comparison with the seed row remains conditional on the adopted illumination.  The seed central depression includes horizon capture and redshift, whereas the shell replaces the excised region by a throat and permits either cross-throat transmission or reflection.  A dark center or a bright inner annulus is therefore a composite consequence of causal class, source support, redshift, and critical compression.  Within $1\leq b_1/M\leq10$ and the common Model-C normalization, the matrix shows that $\bar c_q$ changes the number and placement of returned image features while preserving the principal ordinary peak.  This separation makes the broad-support model especially sensitive to the contralateral optical channel.

\section{Conclusions}\label{sec:conclusions}

We constructed a reflection-asymmetric polymer thin-shell wormhole that isolates the optical action of one-sided quintessence.  The observer and emitting disk occupy a pure-polymer exterior, while the opposite exterior carries the Kiselev-type environment; equal masses, a common polymer scale, the same nonareal angular geometry, and a fixed throat leave the environmental amplitude and exponent as the sources of asymmetry.  The key analytical result is the shell-frame matching of photon frequency and tangential momentum.  It generates a precise map between the intrinsic impact parameters on the two sides and projects the opposite-side photon barrier onto a second critical edge of the physical observer screen.

This matched critical structure organizes the three ray topologies obtained in the calculation.  Observer-side scattering is controlled entirely by the pure-polymer exterior, returned rays cross the throat and turn at the second-side barrier, and terminal rays reach the outer cosmological-type horizon of the finite static patch.  Across the 27 admissible configurations, increasing the dimensionless Kiselev amplitude moved the mapped inner edge inward and widened the returned interval, whereas increasing the dimensionless polymer parameter from $0.40$ to $0.55$ contracted that interval at every sampled pair while preserving it.  The Kiselev exponent produced competing trends because it changes both the throat normalization and the neighborhood of the opposite-side photon sphere.  In parallel, the exact invariance of the observer-side orbit number under changes confined to the second exterior provides a stringent analytical and numerical control of the one-sided construction.

The orbit-number and transfer calculations show how the returned interval becomes a distinct imaging channel.  The two returned orbit-number families accumulate their strongest winding at opposite ends of the allowed interval, directly tracing the two unstable photon barriers.  Ordinary first-, second-, and third-order disk intersections remain fixed when only the opposite-side environment changes, while the returned family responds strongly to the Kiselev parameters.  Within the validated gate, the first new returned disk intersection occurs at third order because the second angular target is crossed before the ray can meet the observer-side disk after recrossing the throat.  This branch emerges near the throat at intermediate environmental amplitude and develops broader radial and screen support as the amplitude increases.  Independent angular scans also reveal the continuation to higher usual and returned orders, confirming a richer near-critical hierarchy beyond the displayed third-order sector.

The intensity profiles and face-on maps translate this geometric hierarchy into emission-dependent image morphology.  The ordinary outer peak remains fixed across the one-sided environmental scan, while returned branches produce additional inner annuli when their disk intersections enter the chosen source support.  For the common radiative slice, the first departure from the black-hole seed occurs at dimensionless Kiselev amplitudes of approximately $0.06895$, $0.04279$, and $0.001116$ for emission beginning at the innermost stable circular orbit, at the photon sphere, and at the throat, respectively.  These markedly separated thresholds quantify how the same cross-throat geometry is filtered by the radial distribution of the emitter.  The broad-support model also exhibits a finite low-impact onset at $b_1/M\simeq2.58330$, where the first disk intersection is created at the throat; its common position across Kiselev amplitudes identifies its observer-side geometric origin.

Taken together, these results establish a controlled optical response to an environment located entirely beyond the throat.  One-sided quintessence shifts the matched inner critical edge, changes the available returned winding, and activates additional image branches while leaving the ordinary observer-side hierarchy invariant.  The separation between invariant local structures and environmentally responsive returned structures provides a clean diagnostic of cross-throat propagation.  The three emission prescriptions further show that image morphology records both the spacetime connection and the radial support of the luminous matter, with the onset amplitudes supplying a quantitative measure of that interplay.

The framework opens several direct extensions.  The local stability formalism developed for reflection-symmetric polymer--quintessence shells can be generalized to the present asymmetric junction, while independent masses, polymer scales, and throat positions would enlarge the matched-critical parameter space.  Higher transfer orders can be incorporated to resolve the full near-critical hierarchy and its integrated flux.  Extending the second-side geometry across the cosmological-type horizon would broaden the causal classification of terminal rays, and inclined or moving emitters, frequency-dependent transfer, absorption, plasma, time variability, and instrument-level image formation would connect the contralateral optical channels to a wider range of astrophysical configurations.

\begin{acknowledgments}
Edson Otoniel would like to thank Fundação Cearense de Apoio ao Desenvolvimento Científico e Tecnológico through grant BP6-0241-00335.01.00/25.
\end{acknowledgments}

\enlargethispage{2\baselineskip}
\bibliography{references}

\end{document}